\documentclass{siamltex}
\usepackage{amsmath,amssymb,amsfonts}   
\usepackage{graphicx,bm}
\usepackage{float}
\usepackage{epsfig}
\usepackage{esint}
 \usepackage{tikz-cd}

\usepackage{bm,braket}

\newcommand{\dis}{\displaystyle}

\newcommand{\calU}{{\mathcal U}}
\newcommand{\calV}{{\mathcal V}}
\newcommand{\calW}{{\mathcal W}}

\newcommand{\calP}{{\mathcal P}}

\newcommand{\calF}{{\mathcal F}}

\newcommand{\calM}{{\mathcal M}}

\newcommand{\calN}{{\mathcal N}}

\renewcommand{\P}{\mathbb{P}}
\newcommand{\PP}{\widetilde{P}}

\newcommand{\x}{\mathbf{x}}

\newcommand{\e}{{\mathrm e}}
\newcommand{\E}{{\mathbb E}}

\newcommand{\calT}{{\mathcal T}}
\renewcommand{\P}{\mathbb P}
\newcommand{\p}{\widetilde{p}}

\begin{document}

\title{Renewal equations for single-particle diffusion in multi-layered media}

\author{Paul C. Bressloff\thanks{Department of Mathematics, University of Utah, Salt Lake City, UT 84112
USA ({\tt bressloff@math.utah.edu})} }

\maketitle

\begin{abstract} 
Diffusion in heterogeneous media partitioned by semi-permeable interfaces has a wide range of applications in the physical and life sciences, ranging from thermal conduction in composite media, gas permeation in soils, diffusion magnetic resonance imaging (dMRI), drug delivery, and intercellular gap junctions. Many of these systems involve three-dimensional (3D) diffusion in an array of parallel planes with homogeneity in the lateral directions, so that they can be reduced to effective one-dimensional (1D) models. 
In this paper we develop a probabilistic model of single-particle diffusion in 1D multi-layered media by constructing a multi-layered version of so-called snapping out Brownian motion (BM).  The latter sews together successive rounds of reflected BM, each of which is restricted to a single layer. Each round of reflected BM is killed when the local time at one end of the layer exceeds an independent, exponentially distributed random variable. (The local time specifies the amount of time a reflected Brownian particle spends in a neighborhood of a boundary.) The particle then immediately resumes reflected BM in the same layer or the layer on the other side of the boundary with equal probability, and the process is iterated
We proceed by constructing a last renewal equation for multi-layered snapping out BM that relates the full probability density to the probability densities of partially reflected BM in each layer. We then show how transfer matrices can be used to solve the Laplace transformed renewal equation, and prove that the renewal equation and corresponding multi-layer diffusion equation are equivalent. We illustrate the theory by analyzing the first passage time (FPT) problem for escape at the exterior boundaries of the domain. Finally, we use the renewal approach to incorporate a generalization of snapping out BM based on the encounter-based method for surface absorption; each round of reflected BM is now killed according to a non-exponential distribution for each local time threshold. This is achieved by considering a corresponding first renewal equation that relates the full probability density to the FPT densities for killing each round of reflected BM. We show that for certain configurations, non-exponential killing leads to an effective time-dependent permeability that is normalizable but heavy-tailed.

\end{abstract}

\section{Introduction}

Diffusion in heterogeneous media partitioned by semi permeable barriers has a wide range of applications in natural and artificial systems. Examples include multilayer electrodes and semi-conductors \cite{Graff04,Diard05,Freger05,Gurevich09}, thermal conduction in composite media \cite{Barbaro88,Grossel98,deMonte00,Lu05}, waste disposal and gas permeation in soils \cite{Yates00,Liu09,Shackelford13}, diffusion magnetic resonance imaging (dMRI) \cite{Tanner78,Callaghan92,Coy94}, drug delivery \cite{Pontrelli07,Todo13}, and intercellular gap junctions \cite{Evans02,Connors04,Good09}. Many of these systems involve three-dimensional (3D) diffusion in an array of parallel planes with homogeneity in the lateral directions, which means that they can be reduced to effective one-dimensional (1D) models. Consequently, there have been a variety of analytical and numerical studies of 1D multilayer diffusion that incorporate methods such as spectral decompositions, Greens functions, and Laplace transforms \cite{Brink85,Ramanan90,Powles92,Dudko04,Hickson09a,Hickson09b,Grebenkov10,Hahn12,Grebenkov14,Carr16,Bressloff16,Moutal19}.

Almost all studies of multilayer diffusion have focused on macroscopic models in which the relevant field is the concentration of diffusing particles. Many of the analytical challenges concern the derivation of time-dependent solutions that characterize short-time transients or threshold crossing events. This requires either carrying out a non-trivial spectral decomposition of the solution and/or inverting a highly complicated Laplace transform. In general, it is necessary to develop some form of approximation scheme or to supplement a semi-analytical solution with numerical computations. As far as we are aware, single-particle diffusion or Brownian motion (BM) in multilayer media has not been investigated to anything like the same extent, with the possible exception of spatially discrete random walks \cite{Kenkre08,Novikov11,Kay22,Alemany22}. On the other hand, a rigorous probabilistic formulation of 1D diffusion through a single semi-permeable barrier has recently been introduced by Lejay \cite{Lejay16} in terms of so-called snapping out BM, see also Refs. \cite{Aho16,Lejay18,Brobowski21}. Snapping out BM sews together successive rounds of reflected BM that are restricted to either $x<0$ or $x>0$ with a semi-permeable barrier at $x=0$. Each round of reflected BM is killed when its local time at $x=0^{\pm}$ exceeds an exponentially distributed random variable with constant rate $\kappa_0$. (Roughly speaking, the local time at $x=0^+$ ($x=0^-$) specifies the amount of time a positively (negatively) reflected Brownian particle spends in a neighborhood of the right-hand (left-hand) side of the barrier \cite{Ito65}.) It then immediately resumes either negatively or positively reflected BM with equal probability, and so on.

We recently reformulated 1D snapping out BM in terms of a renewal equation that relates the full probability density to the probability densities of partially reflected BMs on either side of the barrier \cite{Bressloff23a}.  (The theory of semigroups and resolvent operators were used in Ref. \cite{Lejay16} to derive a corresponding backward equation.) We established the equivalence of the renewal equation with the corresponding single-particle diffusion equation, and showed how to solve the former using a combination of Laplace transforms and Green's function methods. We subsequently extended the theory to bounded domains and higher spatial dimensions \cite{Bressloff23b}. Formulating interfacial diffusion in terms of snapping out BM has at least two useful features. First, it provides a more general probabilistic framework for modeling semi-permeable membranes. For example, each round of partially reflected BM on either side of an interface could be killed according to a non-Markovian process, along analogous lines to encounter-based models of surface absorption \cite{Grebenkov20,Grebenkov22,Bressloff22,Bressloff22a}. That is, partially reflected BM is terminated when its local time at the interface exceeds a random threshold that is not exponentially distributed. As we have shown elsewhere, this leads to a time-dependent permeability that tends to be heavy-tailed \cite{Bressloff23a,Bressloff23b}. Second, numerical simulations of snapping out BM generate sample paths that can be used to obtain approximate solutions of boundary value problems in the presence of  semi-permeable interfaces \cite{Lejay16}.\footnote{An efficient computational schemes for finding solutions to the single-particle diffusion equation in the presence of one or more semi-permeable interfaces has also been developed in terms of underdamped Langevin equations \cite{Farago18,Farago20}. However, this is distinct from snapping out BM, which is an exact single-particle realization of diffusion through an interface in the overdamped limit.}

\begin{figure}[t!]
  \centering
  \includegraphics[width=13cm]{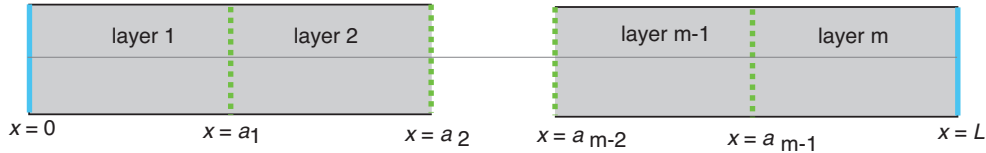}
  \caption{A 1D layered medium consisting of $m$ layers $x\in (a_j,a_{j+1})$, $j=0,1,\ldots m-1$, with $a_0=0$ and $a_m=L$. The interior interfaces at $x=a_j$, $j=1,\ldots m-1$ act as semi-permeable membranes, whereas partially absorbing boundary conditions are imposed on the exterior boundaries at $x=0,L$.}
  \label{fig1}
\end{figure}

In this paper we develop a multi-layered version of snapping out BM and its associated renewal equations for both exponential and non-Markovian killing processes. In particular, we consider a single particle diffusing in a finite interval $[0,L]$ that is partitioned into $m$ subintervals (or layers) $(a_j,a_{j+1})$, $j=0,\ldots,m-1$, with $a_0=0$, $a_m=L$, see Fig. \ref{fig1}. The interior interfaces at $x=a_1,\ldots,a_{m-1}$ are taken to be semi-permeable barriers with constant permeabilities $\kappa_j$, $j=1,\ldots,m-1$, whereas partially reflecting or Robin boundary conditions are imposed at the ends $x=0,L$ with absorption rates $2\kappa_0$ and $2\kappa_l$, respectively. (The factors of 2 are convenient when formulating snapping out BM.) The diffusion coefficient is also heterogeneous with $D(x)=D_j$ for all $x\in (a_{j-1},a_j)$. We begin in section 2 by writing down the multi-layered diffusion equation, which we formally solve using Laplace transforms and an iterative method based on transfer matrices, following along analogous lines to Refs. \cite{Ramanan90,Moutal19}. In section 3, we construct the multi-layered version of snapping out BM and write down the corresponding last renewal equation, which relates the full probability density to the probability densities of partially reflected BM in each layer. We then show how transfer matrices can  be used to solve the Laplace transformed renewal equation, although the details differ significantly from the iterative solution of the diffusion equation. We also prove that the renewal equation and diffusion equation are equivalent. This exploits a subtle feature of partially reflected BM, namely, the Robin boundary condition is modified when the initial position of the particle is on the boundary itself. In section 4 we illustrate the theory by analyzing the first passage time (FPT) problem for the particle to escape from one of the ends of the domain. The FPT statistics can be analyzed in terms of the small-$s$ behavior of the Laplace transformed probability fluxes at the ends $x=0,L$, where $s$ is the Laplace variable. This means that it is sufficient to solve the multi-layer renewal equation in Laplace space, without having to invert the Laplace transformed solution using some form of spectral decomposition, for example. Finally, in section 5,  we use the renewal approach to incorporate a generalization of snapping out BM based on the encounter-based method for surface absorption. This is achieved by considering a corresponding first renewal equation that relates the full probability density to the FPT densities for killing each round of reflected BM.

\section{Single-particle diffusion equation in a 1D layered medium}

Before developing the more general renewal approach for single-particle diffusion in the multi-layer domain of Fig. \ref{fig1}, it is useful to briefly consider the classical formulation in terms of the diffusion equation with constant permeabilities.
Let $\rho_j(x,t)$ denote the probability density of the particle position in the $j$-th layer. For concreteness, we assume that the particle starts in the first layer, that is, $x_0\in [0,a_1]$, although it is straightforward to adapt the analysis to include more general initial conditions, see section 3. (For notational convenience, we drop the explicit dependence of $\rho_j$ on $x_0$.) Single-particle diffusion can be represented by the following piecewise system of partial differential equations (PDEs):
\begin{subequations}
\begin{eqnarray}
\label{FPlayera}
 &&\frac{\partial \rho_j}{\partial t}=D_j\frac{\partial^2\rho_j}{\partial x^2}, \quad x\in (a_{j-1},a_j), \ j=1,\ldots m,\\
 &&\left . D_{j}\frac{\partial \rho_{j}(x,t)}{\partial x}\right |_{x=a_j^{-}}=\left . D_{j+1}\frac{\partial \rho_{j+1}(x,t)}{\partial x}\right |_{x=a_j^{+}}=\kappa_j [\rho_{j+1}(a_j^+,t)-\rho_{j}(a_j^-,t)],\nonumber \\
\label{FPlayerb}
  &&j=1,\ldots m-1,\\
 &&\left . D_1\frac{\partial \rho_1(x,t)}{\partial x}\right |_{x=0}=2\kappa_0 \rho_1(0,t),\quad \left . D_m\frac{\partial \rho_m(x,t)}{\partial x}\right |_{x=L}=-2\kappa_m \rho_m(L,t),
\label{FPlayerc}
\end{eqnarray}
\end{subequations}
together with the initial condition $\rho_j(x,t)=\delta(x-x_0)\delta_{j,1}$. Finally, we denote the composite solution on the domain ${\mathbb G}=\cup_{j=1}^m [a_{j-1}^+,a_j^-]$ by $\rho(x,t)$. 
Laplace transforming equations (\ref{FPlayera})--(\ref{FPlayerc}) gives
\begin{subequations}
\begin{eqnarray}
\label{LTlayera}
 D_j\frac{\partial^2\widetilde{\rho}_j}{\partial x^2}-s\widetilde{\rho}_j&=&-\delta(x-x_0)\delta_{j,1}, \quad x\in (a_{j-1},a_j), \ j=1,\ldots m,\\
 \left . D_{j}\frac{\partial \widetilde{\rho}_{j}(x,s)}{\partial x}\right |_{x=a_j^{-}}&=&\left . D_{j+1}\frac{\partial \widetilde{\rho}_{j+1}(x,s)}{\partial x}\right |_{x=a_j^{+}}=\kappa_j [\widetilde{\rho}_{j+1}(a_j^+,s)-\widetilde{\rho}_{j}(a_j^-,s)],\nonumber \\
\label{LTlayerb}
 &&  j=1,\ldots m-1,\\
 \left . D_1\frac{\partial \widetilde{\rho}_1(x,s)}{\partial x}\right |_{x=0}&=&2\kappa_0 \widetilde{\rho}_1(0,s),\quad \left . D_m\frac{\partial \widetilde{\rho}_m(x,s)}{\partial x}\right |_{x=L}=-2\kappa_m \widetilde{\rho}_m(L,s).
\label{LTlayerc}
\end{eqnarray}
\end{subequations}
Equations (\ref{LTlayera})--(\ref{LTlayerb}) can be solved using transfer matrices along similar lines to Refs. \cite{Ramanan90,Moutal19}. We sketch the steps here.

First, note that for all $1\leq j\leq m$, equation (\ref{LTlayera}) has the general solution
\begin{eqnarray}
 \widetilde{\rho}_j(x,s)=A_j^l(s)\cosh(\sqrt{s/D_j} [x-a_{j-1}])+B_j^l(s)\sinh(\sqrt{s/D_j} [x-a_{j-1}])
\end{eqnarray}
or, equivalently
\begin{eqnarray}
 \widetilde{\rho}_j(x,s)=A_j^r(s)\cosh(\sqrt{s/D_j} [x-a_{j}])+B_j^r(s)\sinh(\sqrt{s/D_j} [x-a_{j}]).
\end{eqnarray}
For $1<j\leq m$, the coefficients $A_j^l,B_j^l$ are related to $A_j^r,B_j^r$ according to
\begin{equation}
 \left (\begin{array}{c} A_j^r \\ B_j^r\end{array}\right )=\calU_j(s)\left (\begin{array}{c} A_j^l \\ B_j^l\end{array}\right ),\quad  \calU_j (s)=\left (\begin{array}{cc} \cosh(\sqrt{s/D_j}L_j )& \sinh(\sqrt{s/D_j}L_j)  \\ \sinh(\sqrt{s/D_j}L_j ) & \cosh(\sqrt{s/D_j}L_j)  \end{array}\right ),
\label{iter0}
\end{equation}
where $L_j=a_j-a_{j-1}$ is the length of the $j$-th layer. The presence of the Dirac delta function for $j=1$ means that the relationship between the coefficients $(A_1^r(s),B_1^r(s))$ and $(A_1^l(s),B_1^l(s))$ is determined by imposing the continuity condition $\widetilde{\rho}_1(x_0^+,s)=\widetilde{\rho}_1(x_0^-,s)$ and the flux discontinuity condition
$\partial_x\widetilde{\rho}_1(x_0^+,s)-\partial_x\widetilde{\rho}_1(x_0^-,s)=-1/D_1$. This yields the result
\begin{equation}
 \left (\begin{array}{c} A_1^r \\ B_1^r\end{array}\right )=\calU_1(s)\left (\begin{array}{c} A_1^l \\ B_1^l\end{array}\right )+\frac{1}{\sqrt{sD_1}}\left (\begin{array}{c} \sinh(\sqrt{s/D_1} [x_0-a_{1}]) \\ -\cosh(\sqrt{s/D_1} [x_0-a_{1}])\end{array}\right ).
\end{equation}
Given the relationships $\widetilde{\rho}_j(a_j,s)=A_j^r(s)$, $ \widetilde{\rho}_j(a_{j-1},s)=A_j^l(s)$, $D_j\partial_x\widetilde{\rho}_j(a_j,s)=\sqrt{sD_j} B_j^r(s)$ and $D_j\partial_x\widetilde{\rho}_j(a_{j-1},s)=\sqrt{sD_j} B_j^l(s)$,
the boundary conditions (\ref{LTlayerb}) can be written in the form
\begin{equation}
\sqrt{sD_j}B_j^r(s)=\sqrt{sD_{j+1}}B_{j+1}^l(s)=\kappa_j[A_{j+1}^l(s)-A_j^r(s)].
\end{equation}
That is, for $1\leq j <m$,
\begin{equation}
 \left (\begin{array}{c} A_{j+1}^l \\ B_{j+1}^l\end{array}\right )=\calV_j(s)\left (\begin{array}{c} A_j^r \\ B_j^r\end{array}\right ),\quad  \calV_j(s) =\left (\begin{array}{cc} 1& \sqrt{sD_j}/\kappa_j \\ & \\ 0 &  \sqrt{D_j/D_{j+1}} \end{array}\right ).
\label{iter}
\end{equation}
Iterating equations (\ref{iter0}) and (\ref{iter}) for $m\geq 2$, we have
\begin{equation}
 \left (\begin{array}{c} A_{m}^r \\ B_{m}^r\end{array}\right )=\calM_m(s) \left (\begin{array}{c} A_1^r\\ B_1^r\end{array}\right ),
\label{iter2}
\end{equation}
with
\begin{equation}
 \calM_2(s)=\calU_2(s)\calV_1(s),\quad \calM_m(s)=\calU_m(s)\left [\prod_{j=2}^{m-1} \calV_{j}(s)\calU_{j}(s)\right ]\calV_1(s) \mbox{ for } m\geq 3 .
\end{equation}
Hence, we have shown how the solution in any layer can be expressed in terms of the two unknown coefficients $A_1^l(s)$ and $B_1^l(s)$. The latter are then determined by imposing the Robin boundary conditions at $x=0,L$:
\begin{eqnarray}
&\sqrt{sD_1}B_1^l(s)=2\kappa_0A_{1}^l(s),\quad \sqrt{sD_m}B_m^r(s)=-2\kappa_mA_{m}^r(s).
\end{eqnarray}

\section{Snapping out BM in a 1D layered medium}

We now develop an alternative formulation of multi-layer diffusion, which is based on a generalization of 1D snapping out BM for a single semi-permeable interface \cite{Lejay16,Bressloff22}. In particular, we construct a renewal equation that relates $\rho(x,t)$ on ${\mathbb G}$ to the probability densities of partially reflected BM in each of the layers $[a_{j-1},a_j]$, $j=1,\ldots,m$. 

\subsection{Single layer with partially reflecting boundaries}

Consider BM in the interval $[a_{j-1},a_j]$ with both ends totally reflecting. Let $X(t)\in [a_{j-1},a_j]$ denote the position of the Brownian particle at time $t$ and introduce the pair of Brownian local times
\begin{subequations}
\begin{eqnarray}
\label{loca}
 \ell_{j-1}^+(t)&=\lim_{h\rightarrow 0} \frac{D_j}{h} \int_0^tH(a_{j-1}+h-X(\tau))d\tau,\\
 \ell_j^-(t)&=\lim_{h\rightarrow 0} \frac{D_j}{h} \int_0^tH(a_j-h-X(\tau))d\tau,
 \label{locb}
\end{eqnarray}
\end{subequations}
where $H$ is the Heaviside function. Note that $\ell_{j-1}^+(t)$ determines the amount of time that the Brownian particle spends in a neighborhood to the right of $x=a_{j-1}$ over the interval $[0,t]$. Similarly, $\ell_{j}^-(t)$ determines the amount of time spent in a neighborhood to the left of $x=a_{j}$. (The inclusion of the factor $D_j$ means that the local times have units of length.) It can be shown that the local times exist and are nondecreasing, continuous function of $t$ \cite{Ito65}. The corresponding stochastic differential equation (SDE) for $X(t)$ is given by the Skorokhod equation
\begin{equation}
dX(t)=\sqrt{2D_j} dW(t)+d\ell_{j-1}^+(t)-d\ell_j^-(t).
\end{equation}
Roughly speaking, each time the particle hits one of the ends it is given an impulsive kick back into the bulk domain. It can be proven that the probability density for particle position evolves according to the single-particle diffusion equation with Neumann boundary conditions at both ends.

\begin{figure}[t!]
  \centering
  \includegraphics[width=10cm]{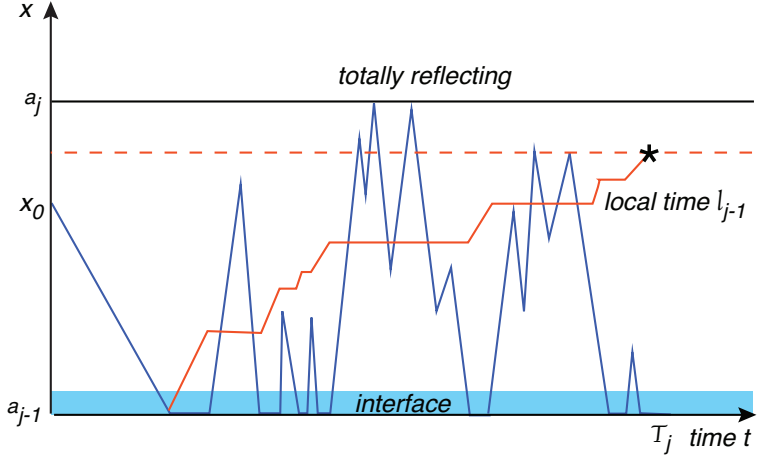}
  \caption{Sketch of a course-grained trajectory of a Brownian particle in the interval $[a_{j-1},a_j]$ with a partially reflecting boundary at $x=a_{j-1}$ and a totally reflecting boundary at $x=a_j$. The particle is absorbed as soon as the time $\ell_{j-1}(t)$ spent in a boundary layer around $x=a_{j-1}$ exceeds an exponentially distribution threshold $\widehat{\ell}_{j-1}$, which occurs at the stopping time $\calT_j$.}
  \label{fig2}
\end{figure}

Partially reflected BM can now be defined by introducing a pair of exponentially distributed independent random local time thresholds $\widehat{\ell}_{j-1}^+$ and $\widehat{\ell}_j^-$ such that
\begin{equation}
 \P[\widehat{\ell}_{j-1}^+>\ell] =\e^{-2\kappa_{j-1}\ell/D_j},\quad \P[\widehat{\ell}_j^->\ell]=\e^{-2\kappa_j\ell/D_j}.
\label{jexp}
\end{equation}
The stochastic process is then killed as soon as one of the local times exceeds its corresponding threshold, which occurs at the stopping time $\calT_j=\min\{\tau_{j}^-,\tau_j^+\}$ with
\begin{equation}
 \tau_{j}^+=\inf\{t>0: \ell_{j-1}^+(t)>\widehat{\ell}_{j-1}^+\}, \quad \tau_j^-=\inf\{t>0: \ell_j^-(t)>\widehat{\ell}_j^-\}.
\label{goat}
\end{equation}
In Fig. \ref{fig2} we illustrate the basic construction using a simplified version of partially reflected BM in which $x=a_{j-1}$ is partially reflecting ($0<\kappa_{j-1}<\infty$) but $x=a_j$ is totally reflecting ($\kappa_j=0$).

It can be shown that the probability density for particle position prior to absorption at one of the ends (see also section 5),
\begin{equation}
 p_j(x,t|x_0)dx=\P[x\leq X(t)<x+dx, t<\calT_j|X_0=x_0],\ x\in [a_{j-1},a_j],
\end{equation}
satisfies the single-particle diffusion equation (Fokker-Planck equation) with Robin boundary conditions at $x=a_{j-1},a_j$ \cite{Freidlin85,Papanicolaou90,Milshtein95,Borodin96,Grebenkov06}:
  \begin{subequations}
  \label{Robin2}
\begin{eqnarray}
 \frac{\partial p_j(x,t|x_0)}{\partial t}&=&D_j\frac{\partial^2 p_j(x,t|x_0)}{\partial x^2}, \quad a_{j-1}<x_0,x<a_j,\\
 D_j\partial_xp_j(a_{j-1},t|x_0)&=&2\kappa_{j-1} p(a_{j-1},t|x_0),\\ 
 D_j\partial_xp_j(a_j,t|x_0)&=&-2\kappa_j p(a_j,t|x_0),
\end{eqnarray}
\end{subequations}
and $p_j(x,0|x_0)=\delta(x-x_0)$. 

It is convenient to Laplace transform with respect to $t$, which gives 
  \begin{subequations}
  \label{RobinLT}
\begin{eqnarray}
 D_j\frac{\partial^2\widetilde{p}_j(x,s|x_0)}{\partial x^2}-s \widetilde{p}_j(x,s|x_0)&=&-\delta(x-x_0),\quad a_{j-1}<x_0,x<a_j\\
 D_j\partial_x\widetilde{p}_j(a_{j-1},s|x_0)&=&2\kappa_{j-1} \widetilde{p}_j(a_{j-1},s|x_0),\\  D_j\partial_x\p(a_j,s|x_0)&=&-2\kappa_j \widetilde{p}_j(a_j,s|x_0).
\end{eqnarray}
\end{subequations}
We can identify $\widetilde{p}_j(x,s|x_0)$ as the Green's function of the modified Helmholtz equation with Robin boundary conditions at $x=a_{j-1},a_j$:
 \begin{eqnarray}
 \widetilde{p}_j(x,s|x_0)= \left \{ \begin{array}{cc} A_j \calF_{j}(x,s) \overline{\calF}_{j}(x_0,s) , & a_{j-1}\leq x\leq x_0\\ & \\
 A_j \calF_{j}(x_0,s) \overline{\calF}_{j}(x,s), & x_0\leq x\leq a_j\end{array}
 \right .
 \label{solVN}
 \end{eqnarray}
where
\begin{subequations}
 \begin{align}
  \calF_{j}(x,s)&=\sqrt{sD_j}\cosh(\sqrt{s/D_j}[x-a_{j-1}])+2\kappa_{j-1} \sinh(\sqrt{s/D_j}[x-a_{j-1}]), \\
   \overline{\calF}_{j}(x,s)&=\sqrt{sD_j}\cosh(\sqrt{s/D_j}[a_{j}-x])+2\kappa_{j} \sinh(\sqrt{s/D_j}[a_{j}-x]), 
 \end{align}
 \begin{eqnarray}
    A_j& =\frac{1}{\sqrt{sD_j}}\frac{1} {2(\kappa_{j-1}+\kappa_j)\sqrt{sD_j}\cosh(\sqrt{s/D_j}L_j)+[sD_j+4\kappa_{j-1}\kappa_j]\sinh(\sqrt{s/D_j}L_j)},\nonumber \\
  \label{A}
  \end{eqnarray}
  \end{subequations}
and $L_j=a_j-a_{j-1}$ is the width of the layer.
It can be checked that the Robin boundary conditions are satisfied at $x=a_{j-1},a_j$ for all $a_{j-1}<x_0<a_j$. However, for $x_0=a_{j-1},a_j$, we have
\begin{subequations}
\begin{eqnarray}
 \label{kola}
D_j\partial_x\widetilde{p}_j(a_{j-1},s|a_{j-1})&=&2\kappa_{j-1}\widetilde{p}(a_{j-1},s|a_{j-1})-1,\\ D_j\partial_x\widetilde{p}_j(a_j,s|a_j)&=&-2\kappa_j\widetilde{p}_j(a_j,s|a_j)+1.
\label{kolb}
 \end{eqnarray}
 \end{subequations}
 In other words,
 \begin{equation}
 \label{lim}
  \lim_{\epsilon \rightarrow 0} \left (\left .\frac{\partial}{\partial x}\right |_{x=a_j}\widetilde{p}_j(x,s|a_j-\epsilon)\right )\neq \left .\frac{\partial}{\partial x}\right |_{x=a_j}\left (\lim_{\epsilon \rightarrow 0} \widetilde{p}_j(x,s|a_j-\epsilon)\right )
 \end{equation}
 etc.
 The modification of the Robin boundary condition when the particle starts at the barrier plays a significant role in establishing the equivalence of snapping out BM with single particle diffusion in a multi-layered medium (see section 3.3).

\subsection{Last renewal equation}

\begin{figure}[t!]
 \raggedleft
  \includegraphics[width=13cm]{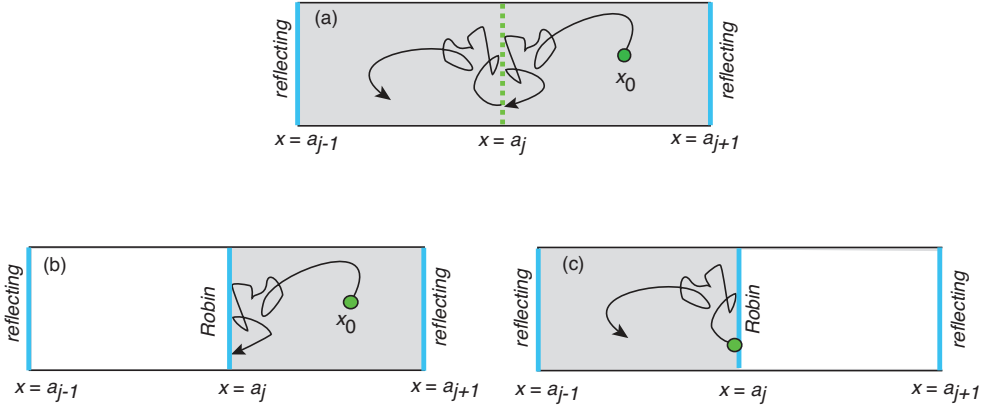}
  \caption{Decomposition of snapping out BM on the interval $[a_{j-1},a_{j+1}]$ with reflecting boundary conditions at the ends and a semi-permeable barrier at $x=a_j$. (a) Diffusion across the interface. (b) Partially reflected BM in $[a_j^+,a_{j+1}]$. (c) Partially reflected BM in $[a_{j-1},a_j^-]$.}
  \label{fig3}
\end{figure}

We now construct snapping out BM in the multi-layered domain shown in Fig. \ref{fig1} by sewing together multiple rounds of reflected BM. For the moment, assume that the exterior  boundaries are totally reflecting. For each interface we introduce a pair of local time $\ell_{j}^{\pm}$ and a corresponding pair of independent exponentially distributed thresholds $\widehat{\ell}_{j}^{\pm}$ with rates $2\kappa_{j}$, $j=1,\ldots,m-1$. Suppose that the particle starts at $x=x_0$ in the first layer. It realizes positively reflected BM until its local time $\ell_1^-(t)$ at $x=a_1$ exceeds the random threshold $\widehat{\ell}_1^-$ with rate $2\kappa_1$. The process immediately restarts as a new reflected BM with probability 1/2 in either $[0,a_1]$ or $[a_1, a_2]$. If the particle is in layer 2, then the reflected BM is stopped as soon as one of the local times $(\ell_1^+(t),\ell_2^-(t))$ exceeds its corresponding threshold. Each time the BM is restarted all local times are reset to zero. Finally, taking the exterior boundaries to be partially reflecting,  we introduce an additional pair of local times, $\ell_0(t),\ell_m(t)$ for the external boundaries at $x=0,L$, and a corresponding pair of exponentially distributed random thresholds $\widehat{\ell}_0,\widehat{\ell}_m$ with rates $2\kappa_0,2\kappa_m$, respectively. The stochastic process is then permanently terminated at the stopping time
\begin{equation}
\calT=\min\{\calT_0,\calT_m\}, \quad \calT_k=\inf\{t>0: \ell_k(t)>\widehat{\ell}_k\},\  k=0,m.
\end{equation}
We illustrate the basic construction in Fig. \ref{fig3} in the simplified case of a single semi-permeable interface at $x=a_j$ and totally reflecting boundaries $x=a_{j-1}$ and $x=a_{j+1}$. The statistics of diffusion across the interface can be captured by sewing together successive rounds of partially reflected BM in the intervals $[a_{j-1},a_{j}^-]$ and $[a_j^+,a_{j+1}]$ with each round killed according to an exponentially distributed local time threshold, and the new domain selected with probability 1/2.

Consider a general initial probability density $\phi(x_0)$ with $x_0\in {\mathbb G}$ and set
\begin{align}
 \rho_j(x,t)&=\int_{\mathbb G}\rho_j(x,t|x_0)\phi(x_0)dx_0,\quad p_j(x,t)=\int_{\mathbb G}p_j(x,t|x_0)\phi(x_0)dx_0.
 \end{align}
Following our previous work on snapping out BM for single semi-permeable interfaces \cite{Bressloff23a,Bressloff23b}, the renewal equation for the $j$-th interior layer, $j=2,\ldots,m-1$, takes the form
\begin{subequations}
 \begin{eqnarray}
   \rho_j(x,t)&= p_j(x,t)+\kappa_{j-1}  \int_0^t p_j(x,\tau|a_{j-1})[\rho_{j-1}(a_{j-1}^-,t-\tau )+\rho_j(a_{j-1}^+,t-\tau )]d\tau \nonumber\\
     &\quad+\kappa_j  \int_0^t p_j(x,\tau|a_{j})[\rho_{j}(a_j^-,t-\tau )+\rho_{j+1}(a_j^+,t-\tau )]d\tau
   \label{renewala}
    \end{eqnarray}
   for all $x\in(a_{j-1}^+,a_j^-)$, with the probability density $p_j(x,\tau|y)$ given by the solution to equations (\ref{Robin2}). The first term $p_j(x,t)$ on the right-hand side of equation (\ref{renewala}) represents all trajectories that reach $x$ at time $t$ without ever being absorbed by the interfaces at $x=a_{j-1}^+,a_j^-$. The first integral on the right-hand side sums over all trajectories that were last absorbed (stopped) at time $t-\tau$ by hitting the interface at $x=a_{j-1}$ from either the left-hand or right-hand side and then switching with probability 1/2 to BM in the $j$-th layer such that it is at position $x \in(a_{j-1}^+,a_j^-)$ at time $t$. Since the particle is not absorbed over the interval $(t-\tau,t]$, the probability of reaching $x$ is $p_j(x,\tau|a_{j-1})$. In addition, the probability that the last stopping event occurred in the interval $(t-\tau,t-\tau+d\tau)$ irrespective of previous events is $2\kappa_{j-1} d\tau$. (We see that the inclusion of the factor 2 in the definition of the permeability cancels the probability factor of 1/2.) The second integral has the corresponding interpretation for trajectories that were last stopped by hitting the interface at $x=a_j$.  In the case of the end layers, we have
 \begin{eqnarray}
  \rho_1(x,t)&=&p_1(x,t)+ \kappa_{1}  \int_0^t p_1(x,\tau|a_1)[\rho_{1}(a_{1}^-,t-\tau )+\rho_2(a_{1}^+,t-\tau )]d\tau,
 \label{renewalb}\\
     \label{renewalc}
  \rho_m(x,t)&=&p_m(x,t)\\
  &&\quad +\kappa_{m-1}  \int_0^t p_m(x,\tau|a_{m-1})[\rho_{m-1}(a_{m-1}^-,t-\tau )+\rho_m(a_{m-1}^+,t-\tau )]d\tau.\nonumber 
    \end{eqnarray}
    \end{subequations}
Note that there is only a single integral contribution in the end layers since only one of the boundaries is semi-permeable. One interesting difference between the renewal equation formulation and the PDE analyzed in section 2 is that the exterior boundary conditions are already incorporated into the solutions $p_1(x,t|x_0)$ and $p_m(x,t|x_0)$, so that they do not have to be imposed separately.
 
Given the fact that the renewal equations (\ref{renewala})--(\ref{renewalc}) are convolutions in time, it is convenient to Laplace transform them by setting $\widetilde{\rho}_j(x,s) =\int_0^{\infty}\e^{-st}\rho_j(x,t)dt$ etc. This gives
 \begin{subequations}
 \begin{align}
 \label{renewal2a}
  \widetilde{\rho}_1(x,s) &= \p_1(x,s)+  \kappa_1 \p_1(x,s|a_1)\Sigma_{1}(s),\, x\in [0^+,a_1^-], \\
   \widetilde{\rho}_j(x,s) &=   \p_j(x,s)+\kappa_{j-1} \p_j(x,s|a_{j-1})\Sigma_{j-1}(s)+\kappa_{j} \p_j(x,s|a_{j})\Sigma_{j}(s),\, x\in [a_{j-1}^+,a_j^-],\nonumber \\
\label{renewal2b}
 &\qquad 1<j <m,\\
\label{renewal2c}
 \widetilde{\rho}_m(x,s) &=  \p_m(x,s)+ \kappa_{m-1} \p_m(x,s|a_{m-1})\Sigma_{m-1}(s),\, x\in [a_{m-1}^+,L^-],
 \end{align}
 \end{subequations}
 where
 \begin{equation}
 \label{Sig}
 \Sigma_{j}(s)=\widetilde{\rho}_j(a_j^-,s )+\widetilde{\rho}_{j+1}(a_j^+,s ) .
 \end{equation}
The functions $\Sigma_j(s)$ can be determined self-consistently by setting $x=a_k^{\pm}$ for $k=1,\ldots,m-1$ and performing various summations. More specifically, substituting equation (\ref{renewal2b}) into the right-hand side of (\ref{Sig}) for $1<j<m$ gives
\begin{subequations}
\begin{align}
 \Sigma_j(s)&=\Sigma_j^p(s)+\kappa_{j-1} \p_j(a_j,s|a_{j-1})\Sigma_{j-1}(s)+\kappa_{j} \p_j(a_j,s|a_{j})\Sigma_{j}(s)\nonumber \\
 &\quad +\kappa_{j} \p_{j+1}(a_j,s|a_{j})\Sigma_{j}(s)+\kappa_{j+1} \p_{j+1}(a_j,s|a_{j+1})\Sigma_{j+1}(s)
\label{Sigja}
\end{align}
for $1<j<m-1$ and 
$\Sigma_j^p(s)\equiv \p_j(a_j,s)+\p_{j+1}(a_{j},s)$.
On the other hand, equations (\ref{renewal2b}) and (\ref{renewal2a}) for $j=2$ implies that
\begin{align}
\Sigma_1(s)&=\Sigma_1^p(s)+\kappa_{1} \p_1(a_1,s|a_{1})\Sigma_{1}(s)\nonumber \\
 &\quad +\kappa_{1} \p_{2}(a_1,s|a_{1})\Sigma_{1}(s)+\kappa_{2} \p_{2}(a_1,s|a_{2})\Sigma_{2}(s),
 \label{Sigjb}
\end{align}
while equations (\ref{renewal2c}) and (\ref{renewal2a}) for $j=m-1$ yields
\begin{align}
\label{Sigjc}
 \Sigma_{m-1}(s)&=\Sigma_{m-1}^p(s)+\kappa_{m-1} \p_m(a_{m-1},s|a_{m-1})\Sigma_{m-1}(s) \\
  & +\kappa_{m-2} \p_{m-1}(a_{m-1},s|a_{m-2})\Sigma_{m-2}(s)+\kappa_{m-1} \p_{m-1}(a_{m-1},s|a_{m-1})\Sigma_{m-1}(s).\nonumber
\end{align}
\end{subequations}

Equations (\ref{Sigja})--(\ref{Sigjc}) can be rewritten in the more compact matrix form
\begin{align}
\sum_{k=1}^{m-1} \Theta_{jk}(s) \Sigma_k(s)=-\Sigma_j^p(s),
\label{sigp}
\end{align}
where ${\bm \Theta}(s)$ is a tridiagonal matrix with non-zero elements
\begin{subequations}
\begin{align}
\Theta_{j,j}(s) &=d_j(s)\equiv\kappa_{j} [\p_{j+1}(a_j,s|a_{j})+\p_j(a_j,s|a_{j})]-1,\ j=1,\ldots m-1,\\
\Theta_{j,j-1}(s)&=  c_{j}(s)\equiv\kappa_{j-1} \p_{j}(a_j,s|a_{j-1}),\quad j=2,\ldots m-1,\\
\Theta_{j,j-1}(s)&= b_{j}(s)\equiv \kappa_{j+1} \p_{j+1}(a_j,s|a_{j+1}),\quad j=1,\ldots,m-2.
\end{align}
\end{subequations}

Assuming that the matrix ${\bm \Theta}(s)$ is invertible, we obtain the formal solution
\begin{equation}
\Sigma_j(s)=-\sum_{k=1}^{m-1} \Theta^{-1}_{jk}(s) \Sigma_k^p(s).
\label{fsigj}
\end{equation}
Substituting into equations (\ref{renewal2a})--(\ref{renewal2c}) gives
\begin{align}
 \widetilde{\rho}_j(x,s) &=   \p_j(x,s)-\sum_{k=1}^{m-1} \left (\kappa_{j-1} \p_j(x,s|a_{j-1})\Theta^{-1}_{j-1,k}(s) +\kappa_{j} \p_j(x,s|a_{j})\Theta^{-1}_{jk}(s) \right )\nonumber \\
 &\times [\p_j(a_j,s)+\p_{j+1}(a_{j+1},s)].
 \label{lastsol}
\end{align}
 
An alternative way to solve for $\Sigma_j(s)$ is to use transfer matrices analogous to the analysis of the PDE in section 2. For simplicity, suppose that the particle starts in the first layer at a point $x_0\in [0,a_1]$ so that 
$\p_j(x,s)=\p_1(x,s|x_0)\delta_{j,1}$.
It follows that equations (\ref{Sigja})--(\ref{Sigjc}) can be rewritten in the iterative form
\begin{equation}
 \left (\begin{array}{c} \Sigma_j\\ \Sigma_{j+1}\end{array}\right )=\calW_j(s)\left (\begin{array}{c} \Sigma_{j-1}\\ \Sigma_j\end{array}\right ),\quad  \calW_j(s) =\left (\begin{array}{cc} 0 & 1  \\ -\frac{\dis c_{j}(s)}{\dis b_j(s)} &-\frac{\dis d_{j}(s)}{\dis b_j(s)} \end{array}\right )
\label{iterR}
\end{equation}
for $1<j<m-1$. In particular,
\begin{equation}
 \left (\begin{array}{c} \Sigma_{m-2}\\ \Sigma_{m-1}\end{array}\right )=\calN(s)\left (\begin{array}{c} \Sigma_{1}\\ \Sigma_2\end{array}\right ),\quad  \calN(s) = \prod_{k=2}^{m-2} \calW_k(s),
 \label{iterR2}
\end{equation}
with, see equation (\ref{Sigjb}), 
\begin{equation}
\Sigma_2(s)=-\frac{1}{b_1(s)}\left (\p_1(a_1,s|x_0)+d_1(s)\Sigma_1(s)\right ).
\end{equation}
Finally, having determined $\Sigma_2,\ldots ,\Sigma_{m-1}$ in terms of $\Sigma_1$, we can calculate $\Sigma_1$ by imposing equation (\ref{Sigjc}), after rewriting it in the more compact form
\begin{equation}
\Sigma_{m-2}(s)=-\frac{d_{m-1}(s)}{c_{m-2}(s)}\Sigma_{m-1}(s).
\end{equation}
We thus obtain the following self-consistency condition for $\Sigma_1$:
\begin{equation}
 \bigg (1, \ \frac{d_{m-1}(s)}{c_{m-2}(s)} \bigg )\calN(s) \left (\begin{array}{c}\Sigma_1(s) \\ -\frac{\dis 1}{\dis b_1(s)}\left (\p_1(a_1,s|x_0)+d_1(s)\Sigma_1(s)\right )\end{array}\right )=0.
\label{iterR3}
\end{equation}

\subsection{Equivalence of the renewal and diffusion equations}

We now have two alternative methods of solution in Laplace space, one based on the diffusion equations (\ref{LTlayera})--(\ref{LTlayerc}) and the other based on the renewal equations (\ref{renewal2a})--(\ref{renewal2c}). Both methods involve transfer matrices that can be iterated to express the solution in the final layer in terms of the solution in the first layer. It is useful to check that the renewal equations (\ref{renewal2a})--(\ref{renewal2c}) are indeed equivalent to the Laplace transformed diffusion equations (\ref{FPlayera})--(\ref{FPlayerc}). (This is simpler than showing that the iterative solutions are equivalent.) Clearly, the composite density $\widetilde{\rho}(x,s)$ satisfies the diffusion equation in the bulk and the exterior boundary conditions, so we only have to check the boundary conditions across the interior interfaces. First, differentiating equations (\ref{renewal2a}) and (\ref{renewal2b}) for $j=2$ with respect to $x$ and setting $x=a_1^{\pm}$ gives
  \begin{subequations}
  \begin{align}
  \label{phoa}
\partial_x \widetilde{\rho}_1(a_1^-,s)&=\partial_x\p_1(a_1,s|x_0)+ \kappa_1\partial_x\p_1(a_1,s|a_1)\Sigma_{1}(s),\\
\label{phob}
\partial_x \widetilde{\rho}_2(a_1^+,s)&=  \kappa_1 
 \partial_x\p_2(a_1,s|a_1)\Sigma_{1}(s) +\kappa_2 \partial_x\p_2(a_1,s|a_2)\Sigma_{2}(s).
  \end{align}
  \end{subequations}
Imposing the Robin boundary condition (\ref{Robin2}) implies that
\[ D_1\partial_x\p_1(a_1,s|x_0)=-2\kappa_1 \p(a_1,s|x_0),\quad D_2\partial_x\p_2(a_1,s|a_2)=2\kappa_1 \p(a_1,s|a_2).\]
 On the other hand, equations (\ref{kola}) and (\ref{kolb}) yield
 \begin{eqnarray*}
  D_1\partial_x\p_1(a_1,s|a_1)&=-2\kappa_1\p(a_1,s|a_1)+1,\quad  D_2\partial_x\p_2(a_1,s|a_1)=2\kappa_1\p_2(a_1,s|a_1)-1.
 \end{eqnarray*}
 Substituting into equations (\ref{phoa}) and (\ref{phob}), we have
   \begin{subequations}
  \begin{align}
    \label{pho2a}
 & D_1\partial_x \widetilde{\rho}_1(a_1^-,s)=-2\kappa_1\p_1(a_1,s|x_0)- \kappa_1 [2\kappa_1\p_1(a_1,s|a_1)-1]\Sigma_{1}(s) ,\\
 & D_2\partial_x \widetilde{\rho}_2(a_1^+,s)=\kappa_1[2\kappa_1\p_2(a_1,s|a_1)-1]\Sigma_1(s)+2\kappa_2
 \kappa_1\p_2(a_1,s|a_2) \Sigma_{2}(s) .
   \label{pho2b}
  \end{align}
  \end{subequations}
Subtracting equations (\ref{pho2a}) and (\ref{pho2b}), and using equation (\ref{Sigjb}) implies that
   \begin{align}
&D_2\partial_x \widetilde{\rho}_2(a_1^+,s)-D_1\partial_x \widetilde{\rho}_1(a_1^-,s) =2\kappa_1\bigg \{\kappa_{1} \p_{2}(a_1,s|a_{1})\Sigma_{1}(s)+\kappa_{2} \p_{2}(a_1,s|a_{2})\Sigma_{2}(s)\nonumber \\
 &\qquad +\p_1(a_1,s|x_0)+\kappa_{1} \p_1(a_1,s|a_{1})\Sigma_{1}(s)-\Sigma_1(s)\bigg \} =0.
 \label{dev1}
  \end{align}
Similarly, adding equations (\ref{pho2a}) and (\ref{pho2b}) gives
  \begin{align}
  &D_2\partial_x \widetilde{\rho}_2(a_1^+,s)+D_1\partial_x \widetilde{\rho}_1(a_1^-,s)] =2\kappa_1\bigg \{\kappa_{1} \p_{2}(a_1,s|a_{1})\Sigma_{1}(s)+\kappa_{2} \p_{2}(a_1,s|a_{2})\Sigma_{2}(s)\nonumber \\
 &\qquad -\p_1(a_1,s|x_0)-\kappa_{1} \p_1(a_1,s|a_{1})\Sigma_{1}(s)\bigg \}.
\end{align}
 On the other hand setting $x=a_1^{\pm}$ in equations (\ref{renewal2a}) and (\ref{renewal2b}) for $j=2$ shows that
 \begin{subequations}
 \begin{align}
 \widetilde{\rho}_1(a_1^-,s) &= \p_1(a_1,s|x_0)+  \kappa_1 \p_1(a_1,s|a_1)\Sigma_{1}(s), \\
   \widetilde{\rho}_2(a_1^+,s) &=  \kappa_{1} \p_2(a_1,s|a_{1})\Sigma_{1}(s)+\kappa_{2} \p_2(a_1,s|a_{2})\Sigma_{2}(s).
 \end{align}
 \end{subequations}
 Hence, we obtain the expected semi-permeable boundary conditions at $x=a_1$,
\begin{eqnarray}
D_2\partial_x \widetilde{\rho}_2(a_1^+,s)=D_1\partial_x \widetilde{\rho}_1(a_1^-,s)=\kappa_1[\widetilde{\rho}_2(a_1^+,s)- \widetilde{\rho}_1(a_1^-,s)].
\label{dev2}
\end{eqnarray}
A similar analysis can be carried out at the other interfaces.

We have thus established the equivalence of the renewal equations (\ref{renewala})--(\ref{renewalc}) and the Laplace transformed diffusion equations (\ref{LTlayera})--(\ref{LTlayerc}). Hence, snapping out BM $X(t)$ on ${\mathbb G}$ is the single-particle realization of the stochastic process whose probability density evolves according to the multi-layer diffusion equation.

 \section{First-passage time problem}

One of the useful features of working in Laplace space is that one can solve various first passage time problems without having to calculate any inverse Laplace transforms. We will illustrate this by considering the escape of the Brownian particle from one of the ends at $x=0,L$. For simplicity, we again assume that the particle starts in the first layer. Let $Q(x_0,t)$ denote the survival probability that a particle starting at $x_0\in (0,a_1)$ has not been absorbed at either end over the interval $[0,t)$. It follows that
\begin{equation}
\label{Q}
Q(x_0,t)=\int_0^L\rho(x,t)dx=\sum_{j=0}^{m-1} \int_{a_j}^{a_{j+1}}\rho_j(x,t)dx.
\end{equation}
(We drop the explicit dependence of $\rho$ and $\rho_j$ on the initial position $x_0$ for notational convenience.) 
Differentiating both sides of equation (\ref{Q}) with respect to $t$ and using equations (\ref{FPlayera})--(\ref{FPlayerc}) shows that
\begin{align}
 \frac{dQ(x_0,t)}{dt}&=\sum_{j=1}^{m} \int_{a_{j-1}}^{a_{j}}\frac{\partial \rho_j(x,t)}{\partial t}dx=\sum_{j=1}^{m} \int_{a_{j-1}}^{a_{j}}D_j\frac{\partial^2 \rho_j(x,t)}{\partial x^2}dx\nonumber \\
 &=\sum_{j=1}^{m} D_j\left [\frac{\partial \rho_j(a_j,t)}{\partial x}-\frac{\partial \rho_j(a_{j-1},t)}{\partial x}\right ]\nonumber \\
 &=D_m \frac{\partial \rho_m(a_m,t)}{\partial x}-D_1\frac{\partial \rho_1(a_{0},t)}{\partial t}\equiv -J_m(x_0,t)-J_0(x_0,t).
\label{Q2}
\end{align}
We have used flux continuity across each interior interface so that the survival probability decreases at a rate equal to the sum of the outward fluxes at the ends $x=0,L$, which are denoted by $J_0$ and $J_L$ respectively. Laplace transforming equation (\ref{Q2}) and imposing the initial condition $Q(x_0,0)=1$ gives
\begin{equation}
\label{QL}
s\widetilde{Q}(x_0,s)-1=-  \widetilde{J}_0(x_0,s)-  \widetilde{J}_L(x_0,s).
\end{equation}
Assuming that $\kappa_0+\kappa_m>0$, the particle is eventually absorbed at one of the ends with probability one, which means that
$\lim_{t\rightarrow \infty}Q(x_0,t)=\lim_{s \rightarrow 0}s\widetilde{Q}(x_0,s) =0$. Hence, $\widetilde{J}_0(x_0,0)+\widetilde{J}_m(x_0,0)=1$. Let $\pi_0(x_0)$ and $\pi_L(x_0)$ denote the splitting probabilities for absorption at $x=0$ and $x=L$, respectively, and denote the corresponding conditional MFPTs by $T_0(x_0)$ and $T_L(x_0)$. It can then be shown that
\begin{equation}
\label{split}
\pi_0(x_0)=\widetilde{J}_0(x_0,0),\quad \pi_L(x_0)=\widetilde{J}_L(x_0,0),
\end{equation}
and
\begin{equation}
\label{Tcon}
 \pi_0(x_0)T_0(x_0)= -\left . \frac{\partial}{\partial s} \widetilde{J}_0(x_0,s)\right |_{s=0}, \quad 
\pi_L(x_0)T_L(x_0)= -\left . \frac{\partial}{\partial s} \widetilde{J}_L(x_0,s)\right |_{s=0}.
\end{equation}
Hence, analyzing the statistics of escape from the domain $[0,L]$ reduces to determining the small-$s$ behavior of the solutions $\partial_x\widetilde{\rho}_1( 0,s)$ and $\partial_x\widetilde{\rho}_m( L,s)$. We will proceed using the renewal equation approach of section 3.

\subsection{Identical layers} A considerable simplification of the iterative equation (\ref{iterR}) occurs in the case of identical layers with $D_j=D$, $\kappa_j=\kappa$ and $a_j=ja$ for all $j=1,\ldots,m$.
The solution (\ref{solVN}) for partially reflected BM is now the same in each layer. That is,
$\p_j(x,s|x_0)=\p(x-(j-1)a,s|x_0-(j-1)a)$ for $x,x_0\in [a_{j-1},a_j]$ with
\begin{eqnarray}
 \widetilde{p}(x,s|x_0)= \left \{ \begin{array}{cc} A \calF(x,s) \overline{\calF}(x_0,s) , & a \leq x\leq x_0\\ & \\
 A \calF(x_0,s) \overline{\calF}(x,s), & x_0\leq x\leq a\end{array}
 \right .,
 \label{solVNs}
 \end{eqnarray}
\begin{subequations}
 \begin{equation}
  \calF(x,s)=\sqrt{sD}\cosh(\sqrt{s/D}[x-a])+2\kappa\sinh(\sqrt{s/D}[x-a]), 
 \end{equation}
 \begin{equation}
  \overline{\calF}(x,s)=\sqrt{sD}\cosh(\sqrt{s/D}[a-x])+2\kappa \sinh(\sqrt{s/D}[a-x]), 
 \end{equation}
 \begin{eqnarray}
    A =\frac{1}{\sqrt{sD}}\frac{1} {4\kappa\sqrt{sD}\cosh(\sqrt{s/D}a)+[sD +4\kappa^2]\sinh(\sqrt{s/D}a)}.  \label{As}
  \end{eqnarray}
  \end{subequations}
In addition equations (\ref{iterR})--(\ref{iterR3}) for identical layers imply that
 \begin{eqnarray}
 \calN(s) =\calW(s)^{m-3},\quad \calW(s)=\left (\begin{array}{cc} 0 & 1  \\ -1 &-g(a,s) \end{array}\right ), \end{eqnarray}
 with
 \begin{eqnarray}
g(y,s)\equiv \frac{2\kappa\p(a,s|y)-1}{\kappa\p(a,s|0)}&=2g_0(y,s) -g_1(s),
 \end{eqnarray}
 where
 \begin{subequations}
 \begin{align}
  g_0(y,s)&\equiv \frac{ \p(a,s|y) }{ \p(a,s|0)}=\frac{\sqrt{sD}\cosh(\sqrt{s/D}y)+2\kappa\sinh(\sqrt{s/D}y)  }{\sqrt{sD}},\\
 g_1(s)&\equiv \frac{1}{\kappa\p(a,s|0)}=\frac{4 \kappa\sqrt{sD}\cosh(\sqrt{s/D}a)+[sD+4\kappa^2]\sinh(\sqrt{s/D}a)}{\kappa\sqrt{sD}}.
 \end{align}
 \end{subequations}
 The matrix $\calW(s)$ can be diagonalized according to
 \begin{equation}
 \label{calWU}
 \calW(s)=\calU \calW_d(s)\calU^{\dagger},\quad \calW_d(s)=\mbox{diag}(\lambda_+(s),\lambda_-(s)),
 \end{equation}
 with
 \begin{equation}
 \lambda_{\pm}(s)=\frac{-g(a,s)\pm \sqrt{g(a,s)^2-4}}{2},\quad \lambda_++\lambda_-=-g,\quad \lambda_+\lambda_-=1,
 \end{equation}
 and
 \begin{equation}
 \calU=\left (\begin{array}{cc} 1 & 1  \\ \lambda_+ &\lambda_- \end{array}\right ),\quad \calU^{\dagger}=\left (\begin{array}{cc} \frac{1}{1-\lambda_+^2} & -\frac{\lambda_+}{1-\lambda_+^2}  \\ \frac{1}{1-\lambda_-^2} &-\frac{\lambda_-}{1-\lambda_-^2} \end{array}\right ),\quad \calU^{\dagger}\calU= \calU\calU^{\dagger}=\left (\begin{array}{cc} 1 & 0 \\ 0 &1 \end{array}\right ).
 \end{equation}
Substituting (\ref{calWU}) into (\ref{iterR2}) and (\ref{iterR3}) gives
\begin{equation}
 \bigg (1, \ g(a,s)  \bigg )\calU(s) \calW_d(s)^{m-3}\calU^{\dagger}(s)\left (\begin{array}{c}\Sigma_1(s) \\ \Sigma_2(s) \end{array}\right )=0,
\label{hiterR}
\end{equation}
and 
\begin{equation}
\Sigma_{m-1}(s)= \bigg (0, \ 1  \bigg )\calU(s) \calW_d(s)^{m-3}\calU^{\dagger}(s)\left (\begin{array}{c}\Sigma_1(s) \\ \Sigma_2(s) \end{array}\right ),
\label{hiterR2}
\end{equation}
with
\begin{equation}
\Sigma_2(s)=-\frac{\dis g_0(x_0,s)}{\dis \kappa }  -g(a,s)\Sigma_1(s).
\end{equation}
In addition, from equations (\ref{renewal2a}) and (\ref{renewal2c}) we have
\begin{align}
 \widetilde{J}_0(x_0,s) &= f(x_0,s)+  \kappa  f(a,s)\Sigma_{1}(s),\quad 
 \widetilde{J}_L(x_0,s)  = \kappa  f(a,s)\Sigma_{m-1}(s),
 \label{Jsplit}
 \end{align}
 where
 \begin{equation}
  D\partial_x\p(0,s|y) = f(y,s) \equiv \frac{2\kappa[\sqrt{sD}\cosh(\sqrt{s/D}[a-y])+2\kappa\sinh(\sqrt{s/D}[a-y])]} {4 \kappa\sqrt{sD}\cosh(\sqrt{s/D}a)+[sD+4\kappa^2]\sinh(\sqrt{s/D}a)},
 \end{equation}
 and $D\partial_x\p(L,s|L-a)=-f(a,s)$.
 
  \begin{figure}[t!]
  \centering
  \includegraphics[width=8cm]{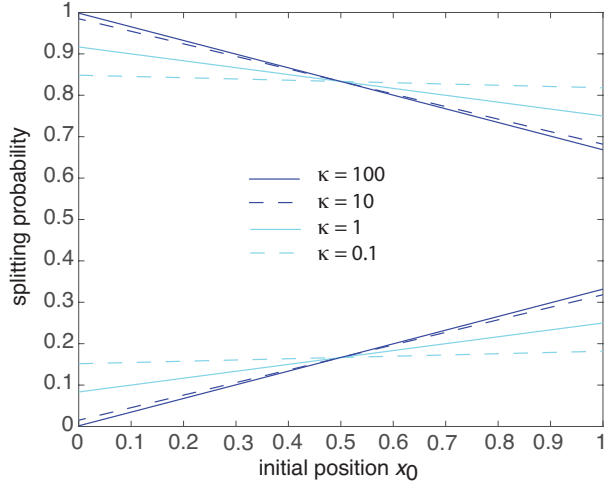}
  \caption{Splitting probabilities for escape from a three-layer, homogeneous medium. Plots of  $\pi_0(x_0)$ and $\pi_L(x_0)$ as a function of $x_0$ for various rates $\kappa$. Other parameters are $D=1$ and $a=1$.}
  \label{fig4}
\end{figure}

For the sake of illustration, consider three layers ($m=3$). Equation (\ref{hiterR}) implies that for $\kappa >0$
\begin{equation}
\Sigma_1(s)=\frac{1}{\dis \kappa }\frac{g(a,s)g_0(x_0,s)}{1-g(a,s)^2},\quad \Sigma_2(s)=-\frac{1}{ \kappa }\frac{g_0(x_0,s)}{1-g(a,s)^2}.
\end{equation}
Using the limits
\begin{align}
  \lim_{s\rightarrow 0}g_0(y,s)&=1+2\kappa y/D,\quad \lim_{s\rightarrow 0}g_1(s)=4(1+\kappa a/D) ,\\
  \lim_{s\rightarrow 0}g(y,s)&=-2(1+2\kappa [a-y]/D),\quad \lim_{s\rightarrow 0}f(y,s)=\frac{(1+2\kappa [a-y]/D)} { 2(1+ \kappa a/D)},
 \end{align}
 we can thus determine the splitting probabilities $\pi_0(x_0)$ and $\pi_L(x_0)$. Example plots of $\pi_0(x_0)$ and $\pi_L(x_0)$ as a function of $x_0\in [0,a]$ are shown in Fig. \ref{fig3} for $a=D=1$. It can be checked that $\pi_0(x_0)+\pi_L(x_0)=1$ for all $x_0$. Moreover, in the limit $\kappa \rightarrow \infty$, we see that $\pi_0(0)\rightarrow 1$ and $\pi_L(0)\rightarrow 0$ as expected. Also note that for $x_0<1/2$ ($x_0>1/2$), $\pi_0(x_0)$ is an increasing (a decreasing) function of $\kappa$.

\subsection{Large number of layers ($m\rightarrow \infty$)}

For a large number of layers ($m\gg 1$) we have 
\begin{equation}
 \calW_d^{m-3}= \left (\begin{array}{cc} \lambda_+^{m-3} & 0  \\ 0 &\lambda_-^{m-3} \end{array}\right )=\lambda_-^{m-3}  \left (\begin{array}{cc} \epsilon & 0  \\ 0 &1\end{array}\right ),\quad \epsilon =\left (\frac{\lambda_+}{\lambda_-}\right )^{m-3}
\end{equation} 
with $|\epsilon |\ll 1$ since $|\lambda_-|>|\lambda_+|$. It follows that
\begin{eqnarray}
\calN(s)=\calU(s) \calW_d(s)^{m-3}\calU^{\dagger}(s)&= \lambda_-(s)^{m-3}\{\calM_0(s)+\epsilon \calM_1(s)\},
\label{pert0}
\end{eqnarray}
where
\begin{eqnarray}
 \calM_0 =\frac{1}{1-\lambda_-^2}\left (\begin{array}{cc} 1 & -\lambda_-  \\ \lambda_- &-\lambda_-^2\end{array}\right ),\quad \calM_1 =\frac{1}{1-\lambda_+^2}\left (\begin{array}{cc} 1 & -\lambda_+ \\ \lambda_+ & -\lambda_+^2\end{array}\right ).
\end{eqnarray}
The next step is to introduce the series expansions
\begin{equation}
\label{pert1}
\Sigma_j(s)=\Sigma_j^{(0)}(s)+\epsilon \Sigma_j^{(1)}(s)+O(\epsilon^2),\quad j=1,2,
\end{equation}
with
\begin{equation}
 \Sigma_2^{(0)}(s) =-\frac{\dis g_0(x_0,s)}{\dis \kappa }  -g(a,s)\Sigma_1^{(0)}(s),\quad \Sigma_2^{(n)}(s) = -g(a,s)\Sigma_1^{(n)}(s)\mbox{ for } n\geq 1.
\end{equation}
Substituting equations (\ref{pert0}) and (\ref{pert1}) into (\ref{hiterR}) and collecting terms in powers of $\epsilon$ gives the $O(1)$ and $O(\epsilon)$ equations
\begin{subequations}
\begin{eqnarray}
 \bigg (1, \ g(a,s)  \bigg )\calM_0(s) \left (\begin{array}{c}\Sigma_1^{(0)}(s) \\ \Sigma_2^{(0)}(s) \end{array}\right )=0,
 \label{hier0}
 \end{eqnarray}
 \begin{eqnarray}
  \bigg (1, \ g(a,s)  \bigg )\left \{\calM_0(s) \left (\begin{array}{c}\Sigma_1^{(1)}(s) \\ \Sigma_2^{(1)}(s) \end{array}\right ) +\calM_1(s) \left (\begin{array}{c}\Sigma_1^{(0)}(s) \\ \Sigma_2^{(0)}(s) \end{array}\right )\right \}
 =0.
 \label{hier1}
 \end{eqnarray}
 \end{subequations}
 Equation (\ref{hier0}) has the solution
 \begin{equation}
 \Sigma_1^{(0)}(s)=-\frac{\lambda_-(s)g_0(x_0,s)}{\kappa (1+g(a,s)\lambda_-(s))}=\frac{g_0(x_0,s)}{\kappa \lambda_-(s)},
 \end{equation}
 so that
 \begin{equation}
  \Sigma_1^{(1)}(s)=\frac{\lambda_+(s)^4}{\lambda_-(s)^4}\frac{1-\lambda_+(s)^2}{1-\lambda_-(s)^2}\left (\Sigma_1^{(0)}-\frac{g_0(x_0,s)}{\kappa \lambda_+(s)}\right ).
  \end{equation}
  Finally,
\begin{align}
 \Sigma_{m-1}(s)&=(0, \ 1) \calN(s) \left (\begin{array}{c}\Sigma_1(s) \\ \Sigma_2(s) \end{array}\right )
\\
 &=\lambda_-(s)^{m-3} (0,\ 1) \{\calM_0(s)+\epsilon \calM_1(s)\} \left (\begin{array}{c}\Sigma_1^{(0)}(s) +\epsilon \Sigma_1^{(1)}(s)+O(\epsilon^2) \\ \Sigma_2^{(0)}(s) +\epsilon \Sigma_1^{(1)}(s) +O(\epsilon^2)\end{array}\right )  \nonumber \\
 &=\lambda_+(s)^{m-3} (0,\ 1)  \left \{\calM_0(s) \left (\begin{array}{c}\Sigma_1^{(1)}(s) \\ \Sigma_2^{(1)}(s) \end{array}\right ) +\calM_1(s) \left (\begin{array}{c}\Sigma_1^{(0)}(s) \\ \Sigma_2^{(0)}(s) \end{array}\right )+O(\epsilon)\right \}.\nonumber
\end{align}
We have used the fact the $O(1)$ solution $(\Sigma_1^{(0)},\Sigma_2^{(0)})^{\top}$ is actually a null-vector of the matrix $\calM_0$ so the leading contribution to $\Sigma_{m-1}(s)$ is proportional to $\epsilon \lambda_-(s)^{m-3}=\lambda_+(s)^{m-3}$. Hence, $\Sigma_{m-1}(s)\rightarrow 0$ as $m\rightarrow \infty$ due to the fact that $|\lambda_+(s)| < 1$ for all $s$. Equations (\ref{split}) and (\ref{Jsplit}) then imply that $\pi_m(\x_0)\rightarrow 0$ as $m \rightarrow \infty$, with the rate of decay determined by $\lambda_+(0)^{m-3}$.

\section{Generalized model of multi-layer diffusion}

The analysis of the FPT problem in section 4 could also have been carried out using the solution of the diffusion equation constructed in section 2. However, one advantage of the renewal approach is that it is based on snapping out BM, which can be used to generate sample paths of single-particle diffusion in a multi-layer medium. Rather than exploring numerical aspects here, we consider another advantage of the renewal approach, namely, it supports a more general model of semi-permeable membranes. This is based on an extension of snapping out BM that modifies the rule for killing each round of reflected BM within a layer. We proceed by applying the encounter-based model of absorption \cite{Grebenkov20,Grebenkov22,Bressloff22,Bressloff22a} to reflected BM in each of the layers separately. 

\subsection{Local time propagator for a single layer}

As we mentioned in section 3.1, partially reflected BM in an interval can be implemented by introducing exponentially distributed local time thresholds at either end of the interval, which then determine when reflected BM is killed. Here we generalize the killing mechanism.
Given the local times (\ref{loca}) and (\ref{locb}) of the $j$-th layer with totally reflecting boundaries, the local time propagator is defined according to \cite{Grebenkov20}
\begin{align}
&P_j(x,\ell,\ell^{\prime},t|x_0)dx\, d\ell\, \ell^{\prime} \nonumber \\
 &=\P[x<X(t)<x+dx,\ell<\ell_{j-1}^+<\ell+d\ell,
\ell^{\prime}<\ell_{j}^-<\ell^{\prime}+d\ell^{\prime}|X(0)=x_0].
\end{align}
Next, for each interface we introduce a pair of  independent identically distributed random local time thresholds $\widehat{\ell}_{j}^{\pm}$ such that
$ \P[\widehat{\ell}_{j}^{\pm}>\ell]\equiv \Psi_{j}^{\pm}(\ell)$.
The special case of exponential distributions is given by equations (\ref{jexp}).
The stochastic process in the $j$-th layer is then killed as soon as one of the local times $ {\ell}_{j-1}^{+}$ and $ {\ell}_j^-$ exceeds its corresponding threshold, which occurs at the FPT time $\calT_j=\min\{\tau_{j}^+,\tau_j^-\}$, see equation (\ref{goat}).
Since the corresponding local time thresholds $\widehat{\ell}_{j-1}^{+}$ and $\widehat{\ell}_j^-$ are statistically independent, the relationship between the resulting probability density $p_j(x,t|x_0)$ for partially reflected BM in the $j$-th layer and $P_j(x,\ell_1,\ell_2,t|x_0)$ can be established as follows:
\begin{align*}
 & p_j(x,t|x_0)dx=\P[X(t) \in (x,x+dx), \ t < {\mathcal T}_j|X_0=x_0]\\
 &=\P[X(t) \in (x,x+dx), \ \ell_{j-1}^+(t)< \widehat{\ell}_{j-1}^+,\ \ell_{j}^-(t)< \widehat{\ell}_{j}^-|X_0=x_0]\\
 &=\int_0^{\infty} d\ell \psi_{j-1}^+(\ell) \int_0^{\infty} d\ell^{\prime} \psi_j^-(\ell^{\prime}) \P[X(t) \in (x,x+dx), \ \ell_{j-1}^+ < \ell, \ \ell_{j}^- <\ell^{\prime}|X_0=x_0]\\
 &=\int_0^{\infty} d\ell \psi_{j-1}^+(\ell) \int_0^{\infty} d\ell^{\prime} \psi_j^-(\ell^{\prime})\int_0^{\ell} d\hat{\ell}  \int_0^{\ell^{\prime}}d\hat{\ell}^{\prime}[P_j(x,\hat{\ell} ,\hat{\ell} ',t|x_0)dx].
\end{align*}
We have also introduced the probability densities $\psi_j^{\pm}(\ell)=-\partial_{\ell}\Psi_j^{\pm}(\ell)$.
Reversing the orders of integration yields the result
\begin{equation}
\label{mbob}
p_j(x,t|x_0)=\int_0^{\infty}d\ell\Psi_{j-1}^+(\ell) \int_0^{\infty} d\ell^{\prime}\Psi_j^-(\ell^{\prime}) P_j(x,\ell,\ell^{\prime},t|x_0).
\end{equation}

An evolution equation for the local time propagator can be derived as follows \cite{Bressloff22,Bressloff22a}. Since the local times only change at the boundaries $x=a_{j-1},a_j$, the propagator satisfies the diffusion equation in the bulk of the domain
 \begin{eqnarray} 
   \label{JPCK1}
   \frac{\partial P_j}{\partial t}  = D_j\frac{\partial^2 P_j}{\partial x^2},\ x\in (a_{j-1},a_j).
     \end{eqnarray}
The nontrivial step is determining the boundary conditions at $x=a_{j-1},a_j$. Here we give a heuristic derivation based on a boundary layer construction. For concreteness, consider the left-hand boundary layer $[a_{j-1},a_{j-1}+h]$ and define
\begin{eqnarray}
\ell_{j-1}^h(t)=\frac{D_j}{h} \int_0^t\left [\int_{0}^h\delta(X_{t'}-x)dx\right ]dt' .
\end{eqnarray}
By definition, $h\ell_{j-1}^h(t)/D_j$ is the residence or occupation time of the process $X(t)$ in the boundary layer up to time $t$.
Although the width $h$ and the residence time in the boundary layer vanish in the limit $h\rightarrow 0$, the rescaling by $1/h$ ensures that $\lim_{h\rightarrow 0} \ell_{j-1}^h(t)=\ell_{j-1}^+(t)$.
Moreover, from conservation of probability, the flux into the boundary layer over the residence time $h\delta \ell/D_j$ generates a corresponding shift in the probability $P_j$ within the boundary layer from $\ell\rightarrow \ell +\delta \ell$. That is, for $\ell>0$,
\begin{eqnarray*}
 -J_j(a_{j-1}+h,\ell,\ell^{\prime},t|x_0)h\delta \ell =[P_j(a_{j-1},\ell+\delta\ell,\ell^{\prime},t|x_0)-P_j(a_{j-1},\ell,\ell^{\prime},t|x_0)]h ,
\end{eqnarray*}
where $J_j(x,\ell,\ell^{\prime},t|x_0)=-D\partial_x P_j(x,\ell,\ell^{\prime},t|x_0)$.
Dividing through by $h\delta \ell$ and taking the limits $h\rightarrow 0$ and $\delta \ell \rightarrow 0$ yields 
 \[-J_j(a_{j-1},\ell,\ell^{\prime},t|x_0)=\partial_{\ell}P_j(a_{j-1},\ell,\ell^{\prime},t|x_0),\ \ell >0.\]
Moreover, when $\ell=0$ the probability flux $J_j(a_{j-1},0,\ell^{\prime},t|x_0)$ is identical to that of a Brownian particle with a totally absorbing boundary at $x=a_{j-1}$, which we denote by $J_{j,\infty}(a_{j-1},\ell^{\prime},t|x_0)$. In addition, it can be shown that $P_j(a_{j-1},0,\ell^{\prime},t|x_0)=-J_{j,\infty}(a_{j-1},\ell^{\prime},t|x_0)$. Applying a similar argument at the end $x=a_j$, we obtain the pair of boundary conditions
\begin{subequations}
\begin{align}
\label{BCLa}
 D\partial_x P_j(a_{j-1},\ell,\ell^{\prime},t|x_0)&=-P_j(a_{j-1},0,\ell^{\prime},t|x_0)\delta(\ell )+\frac{\partial P_j(a_{j-1},\ell,\ell^{\prime},t|x_0)}{\partial \ell},\\
 -D\partial_x P_j(a_{j},\ell,\ell^{\prime},t|x_0)&=-P_j(a_{j},\ell,0,t|x_0)\delta(\ell^{\prime})+\frac{\partial P_j(a_{j},\ell,\ell^{\prime},t|x_0)}{\partial \ell^{\prime}}.
\label{BCLb}
\end{align}
\end{subequations}

The crucial step in the encounter-based approach is to note that for exponentially distributed local time thresholds, see equation (\ref{jexp}), the right-hand side of equation (\ref{mbob}) reduces to a double Laplace transform of the local time propagator:
\begin{equation}
\label{mbobLT}
 p_j(x,t|x_0)=\calP_j(x,z_{j-1}^+,z_j^-,t|x_0),\quad z_{j-1}^+=\frac{2\kappa_{j-1}}{D_j},\quad z_j^-=\frac{2\kappa_{j}}{D_j},
\end{equation}
with
\begin{equation}
 \calP_j(x,z,z',t|x_0)\equiv \int_0^{\infty}d\ell \e^{-z \ell }\int_0^{\infty} d\ell^{\prime}\e^{-z'\ell^{\prime}} P_j(x,\ell,\ell^{\prime},t|x_0).
\end{equation}
Laplace transforming the propagator boundary conditions (\ref{BCLa}) and (\ref{BCLb}) then shows that the probability density $p_j$ of equation (\ref{mbobLT}) is the solution to the Robin BVP given by equations (\ref{Robin2}a) and  (\ref{Robin2}b).
Hence, the probability density of partially reflected BM in the $j$-th layer is equivalent to the doubly Laplace transformed local time propagator with the pair of Laplace variables $z_{j-1}^+$ and $z_j^-$. Assuming that the Laplace transforms can be inverted, we can then incorporate non-exponential probability distributions $\Psi_{j-1}^+(\ell)$ and $\Psi_{j}^-(\ell^{\prime})$ such that the corresponding marginal density is now
  \begin{equation}
  \label{oo}
   p_j (x,t|x_0)=\int_0^{\infty}d\ell\Psi_{j-1}^+(\ell) \int_0^{\infty} d\ell^{\prime}\Psi_j^-(\ell^{\prime}){\mathcal L}_{\ell}^{-1}{\mathcal L}_{\ell^{\prime}}^{-1}\calP_j(x,z,z',t|x_0),
  \end{equation}
  where ${\mathcal L}^{-1}$ denotes the inverse Laplace transform.
  One major difference from the exponential case is that the stochastic process $X(t)$ is no longer Markovian. 
  
  \subsection{Killing time densities}
  
  In order to sew together successive rounds of reflected BM in the case of general distributions $\Psi_j$ we will need the conditional FPT densities $f_{j-1}^{+}(x_0,t)$ and $f_{j}^{-}(x_0,t)$ for partially reflected BM in the $j$-th layer to be killed at the ends $x=a_{j-1}$ and $x=a_j$, respectively. The corresponding conditional killing times  were defined in equation (\ref{goat}). The FPT densities are given by the outward probability fluxes at the two ends:
  \begin{equation}
  \label{fpm}
  f_{j-1}^+(x_0,t)=D_j \partial_x p_j (a_{j-1},t|x_0),\quad f_j^-(x_0,t)=-D_j \partial_xp_j (a_{j},t|x_0).
  \end{equation}
 As in previous sections, it is convenient to Laplace transform with respect to $t$.
  Laplace transforming equation (\ref{oo}) and using the Green's function (\ref{solVN}) gives
   \begin{equation}
  \label{zoo}
   \p_j (x,s|x_0)=\int_0^{\infty}d\ell\Psi_{j-1}^+(\ell) \int_0^{\infty} d\ell^{\prime}\Psi_j^-(\ell^{\prime}){\mathcal L}_{\ell}^{-1}{\mathcal L}_{\ell^{\prime}}^{-1}\widetilde{\calP}_j(x,z,z',s|x_0),
  \end{equation}
  where
   \begin{eqnarray}
 \widetilde{\calP}_j(x,z,z',s|x_0)= \left \{ \begin{array}{cc} A_j(z,z',s) \calF_{j}(x,z,s) \overline{\calF}_{j}(x_0,z',s) , & a_{j-1}\leq x\leq x_0,\\ & \\
 A_j(z,z',s) \calF_{j}(x_0,z,s) \overline{\calF}_{j}(x,z',s), & x_0\leq x\leq a_j,\end{array}
 \right .
 \label{zsolVN}
 \end{eqnarray}
with
\begin{subequations}
 \begin{align}
  \calF_{j}(x,z,s)&=\sqrt{s/D_j}\cosh(\sqrt{s/D_j}[x-a_{j-1}])+z\sinh(\sqrt{s/D_j}[x-a_{j-1}]), \\
   \overline{\calF}_{j}(x,z',s)&=\sqrt{s/D_j}\cosh(\sqrt{s/D_j}[a_{j}-x])+z' \sinh(\sqrt{s/D_j}[a_{j}-x]), \\
    A_j& =\frac{1}{\sqrt{sD_j}}\frac{1} {(z+z')\sqrt{s/D_j}\cosh(\sqrt{s/D_j}L_j)+[s/D_j+zz']\sinh(\sqrt{s/D_j}L_j)}.\nonumber \\
  \label{zA}
  \end{align}
  \end{subequations}
  Since $\PP_j(x,z,z,s|x_0)$ satisfies the Robin boundary conditions
  \begin{align*}
  D_j\partial_x\widetilde{\calP}_j(a_{j-1},z,z',s|x_0)=D_jz\widetilde{\calP}_j(a_{j-1},z,z',s|x_0),\\ D_j\partial_x\widetilde{\calP}_j(a_{j},z,z',s|x_0)=-D_jz'\widetilde{\calP}_j(a_{j},z,z',s|x_0),
  \end{align*}
   it follows that
  \begin{align}
&\widetilde{f}^+_{j-1}(x_0,s) \equiv D_j\partial_x\p_j(a_{j-1},s|x_0)\nonumber \\
 &=D_j\int_0^{\infty}d\ell\Psi_{j-1}^+(\ell) \int_0^{\infty} d\ell^{\prime}\Psi_j^-(\ell^{\prime})\left [\partial_{\ell}\widetilde{P}_j(a_{j-1},\ell,\ell',s|x_0)+\widetilde{P}_j(a_{j-1},0,\ell',s|x_0)\right ]\nonumber \\
  &=D_j\int_0^{\infty}d\ell\psi_{j-1}^+(\ell) \int_0^{\infty} d\ell^{\prime}\Psi_j^-(\ell^{\prime}) \widetilde{P}_j(a_{j-1},\ell,\ell',s|x_0).
  \label{fm}
  \end{align}
  Similarly,
  \begin{align}
  \label{fp}
 \widetilde{f}^-_j(x_0,s) \equiv -D_j\partial_x\p_j(a_{j},s|x_0)=D_j\int_0^{\infty}d\ell\Psi_{j-1}^+(\ell) \int_0^{\infty} d\ell^{\prime}\psi_j^-(\ell^{\prime}) \widetilde{P}_j(a_j,\ell,\ell',s|x_0).
  \end{align}

Evaluation of the FPT densities reduces to the problem of calculating the propagator $\widetilde{P}_j(a_k,\ell,\ell',s|x_0)$ by inverting the double Laplace transform $\widetilde{\calP}_j(a_k,z,z',s|x_0)$ with respect to $z$ and $z'$, $k=j-1,j$, and then evaluating the double integrals in equations (\ref{fm}) and (\ref{fp}). In general, this is a non-trivial calculation. However, a major simplification occurs if we take one of the densities $\Psi_{j-1}^+$ or $\Psi_j^-$ to be an exponential. First suppose that $\Psi_{j-1}^+(\ell)=\e^{-2\kappa_{j-1}\ell/D_j}$. We then have a Robin boundary condition at $x=a_{j-1}$,
\begin{equation}
\widetilde{f}^+_{j-1}(x_0,s) =2\kappa_{j-1} \p_j(a_{j-1},s|x_0),
\end{equation}
whereas
\begin{align}
\widetilde{f}^-_j(x_0,s)=D_j\int_0^{\infty} d\ell^{\prime}\psi_j^-(\ell^{\prime}) \widetilde{P}_j(a_j,z_{j-1}^+,\ell',s|x_0).
\end{align}
From equation (\ref{zoo}) we find that
\begin{align}
\widetilde{\calP}_j(a_j,z_j,z',s|x_0)&=\frac{1}{D_j} \frac{\Lambda_j(x_0,s)}{z'+h_j(s)},
\end{align}
where
\begin{equation}
\label{Lam}
\Lambda_j(x_0,s)=\frac{\sqrt{s/D_j}\cosh(\sqrt{s/D_j}[x_0-a_{j-1}])+z_{j-1}^+\sinh(\sqrt{s/D_j}[x_0-a_{j-1}])}{ \sqrt{s/D_j}\cosh(\sqrt{s/D_j}L_j)+z_{j-1}^+\sinh(\sqrt{s/D_j}L_j},
\end{equation}
and
\begin{equation}
\label{hj}
h_j(s)=\sqrt{s/D_j}\frac{\sqrt{s/D_j} \tanh(\sqrt{s/D_j}L_j)+z_{j-1}^+}{\sqrt{s/D_j}+z_{j-1}^+\tanh(\sqrt{s/D_j}L_j )}.
\end{equation}
Inverting the Laplace transform with respect to $z'$ then gives
\begin{align}
 \widetilde{P}_j(a_j,z_{j-1}^+,\ell',s|x_0)=D_j^{-1}\Lambda_j(x_0,s)\e^{-h_j(s)\ell'}
 \end{align}
 and, hence,
 \begin{align}
\widetilde{f}^-_j(x_0,s)=\Lambda_j(x_0,s) \widetilde{\psi}_j^-(h_j(s)) .
\end{align}
On the other hand,
\begin{equation}
\p_j(a_j,s|x_0)=D_j^{-1}\Lambda_j(x_0,s) \widetilde{\Psi}_j^-(h_j(s)).
\end{equation}
We thus obtain the following boundary condition at $x=a_j$:
\begin{equation}
\widetilde{f}^-_j(x_0,s)=\widetilde{K}_j^-(s) \p_j(a_j,s|x_0),\quad \widetilde{K}^-_j(s)=\frac{D_j\widetilde{\psi}^-_j(h_j(s))}{\widetilde{\Psi}^-_j(h_j(s))}.
\end{equation}
Finally, using the convolution theorem, the boundary condition at $x=a_j$ in the time domain takes the form
 \begin{eqnarray}
D_j \partial_x p_j(a_j,t|x_0)=-\int_0^t K_j^-(\tau)p_j(a_j,t-\tau|x_0)d\tau.
\end{eqnarray}
That is, in the case of a non-Markovian density for killing partially reflected BM at one end of an interval, the corresponding boundary condition involves an effective time-dependent absorption rate $K_j^-(t)$, which acts as a memory kernel. 

Now suppose that $\Psi_{j}^-(\ell)=\e^{-2\kappa_{j}\ell/D_j}$ so that 
\begin{align}
\widetilde{f}^-_{j}(x_0,s) =2\kappa_{j} \p_j(a_{j},s|x_0),\
\widetilde{f}^+_{j-1}(x_0,s)=D_j\int_0^{\infty} d\ell \psi_{j-1}^+(\ell) \widetilde{P}_j(a_{j-1},\ell, z_j^-,s|x_0).
\end{align}
From equation (\ref{zoo}) we have
\begin{align}
\widetilde{\calP}_j(a_j,z,z_j^-,s|x_0)&=\frac{1}{D_j} \frac{\overline{\Lambda}_j(x_0,s)}{z+\overline{h}_j(s)},
\end{align}
where
\begin{equation}
\label{Lam2}
\overline{\Lambda}_j(x_0,s)=\frac{\sqrt{s/D_j}\cosh(\sqrt{s/D_j}[a_j-x_0])+z_{j}^-\sinh(\sqrt{s/D_j}[a_j-x_0])}{ \sqrt{s/D_j}\cosh(\sqrt{s/D_j}L_j)+z_{j}^-\sinh(\sqrt{s/D_j}L_j},
\end{equation}
and
\begin{equation}
\label{hj2}
\overline{h}_j(s)=\sqrt{s/D_j}\frac{\sqrt{s/D_j} \tanh(\sqrt{s/D_j}L_j)+z_{j}^-}{\sqrt{s/D_j}+z_{j}^-\tanh(\sqrt{s/D_j}L_j )}.
\end{equation}
Using identical arguments to the previous case we find that the boundary condition at $x=a_{j-1}$ is
\begin{equation}
\widetilde{f}^+_{j-1}(x_0,s)=\widetilde{K}_{j-1}^+(s) \p_{j}(a_{j-1},s|x_0),\quad \widetilde{K}^+_{j-1}(s)=\frac{D_j\widetilde{\psi}^+_{j-1}(\overline{h}_j(s))}{\widetilde{\Psi}^+_{j-1}(\overline{h}_j(s))}.
\end{equation}

\subsection{Generalized snapping out BM and the first renewal equation}

We now define a generalized snapping out BM by sewing together successive rounds of reflected BM along identical lines to section 3.2, except that now each round is killed according to the general process introduced in section 5.1. (For simplicity, we assume that the exterior boundaries at $x=0,L$ are totally reflecting.) Although each round of partially reflected Brownian motion is non-Markovian, all history is lost following absorption and restart so that we can construct a renewal equation. However, it is now more convenient to use a first rather than a last renewal equation. Again we consider a general probability density $\phi(x_0)$ of initial conditions $x_0\in {\mathbb G}$.

 Let $f_{j-1}^{+}(t)$ and $f_{j}^{-}(t)$ denote the conditional FPT densities for partially reflected BM in the $j$-th layer to be killed at the end $x=a_{j-1}$ and $x=a_j$, respectively, in the case of a general initial distribution $\phi(x_0)$. It follows that 
\begin{subequations}
\label{fptj}
 \begin{align}
  f_{j-1}^+(t)&=\int_{a_{j-1}}^{a_j} f_{j-1}^+(x_0,t)\phi(x_0)dx_0=D_j\int_{a_{j-1}}^{a_j}\partial_x p_j(a_{j-1},t|x_0)\phi(x_0)dx_0,\\  
 f_{j}^-(t)&=\int_{a_{j-1}}^{a_j} f_{j}^-(x_0,t)\phi(x_0)dx_0=-D_j\int_{a_{j-1}}^{a_j} \partial_x p_j(a_{j},t|x_0)\phi(x_0) dx_0.
 \end{align}
 \end{subequations}
 with $f_{j-1}^{+}(x_0,t)$ and $f_{j}^{-}(x_0,t)$ defined in equations (\ref{fpm}).
 We also set $f_1^+(t)\equiv 0$ and $f_m^-(t)\equiv 0$. Generalizing previous work \cite{Bressloff23a,Bressloff23b}, the first renewal equation in the $j$-th layer, $1\leq j\leq m$, takes the form
 \begin{align}
  \label{Frenewal}
  \rho_j(x,t)&\equiv \int_{\mathbb G}\rho_j(x,t|x_0)\phi(x_0)dx_0\\
  &=p_j(x,t)+\frac{1}{2}\sum_{k=1}^{m-1}\int_0^t [\rho_{j}(x,t-\tau |a_{k}^-)+\rho_j(x,t-\tau|a_{k}^+ )][f_{k}^-(\tau)+f_{k}^+(\tau)]d\tau \nonumber 
   \end{align}
   for $x \in (a_{j-1},a_j)$ and
   \begin{equation}
   p_j(x,t)=\int_{a_{j-1}}^{a_j}p_j(x,t|x_0)\phi(x_0)dx_0.
   \end{equation}
 The first term on the right-hand side of equation (\ref{Frenewal}) represents all sample trajectories that start in the $k$-th layer and have not been absorbed at the ends $x=a_{k-1},a_k$ up to time $t$. The integral term represents all trajectories that were first absorbed (stopped) at a semi-permeable interface at time $\tau$ and then switched to either positively or negatively reflected BM state with probability 1/2, after which an arbitrary number of switches can occur before reaching $x\in (a_{j-1},a_j)$ at time $t$. The probability that the first stopping event occurred at the $k$-th interface in the interval $(\tau,\tau+d\tau)$ is $[f_{k}^+(\tau)+f_{k}^-(\tau)]d\tau$. Laplace transforming the renewal equation (\ref{Frenewal}) with respect to time $t$ gives
 \begin{align}
  \label{Frenewal2}
  \widetilde{\rho}_j(x,s) &=  \p_j(x,s)   +\frac{1}{2}\sum_{k=1}^{m-1}[\widetilde{\rho}_j(x,s |a_{k}^-)+\widetilde{\rho}_j(x,s |a_{k}^+) ][\widetilde{f}_{k}^-(s)
+\widetilde{f}_{k}^+(s)] .
 \end{align}
In order to determine the factors 
\begin{align}
\Sigma_{jk}(x,s)&=  \widetilde{\rho}_j(x,s |a_k^-)+\widetilde{\rho}_j(x,s |a_k^+) , \quad 1\leq k<m, 
\label{initial}
 \end{align}
we  substitute  into equation (\ref{Frenewal2}) the initial density
$ \phi(x_0)=\frac{1}{2}[\delta(x_0-a_k^-) +\delta(x_0-a_k^+)]. $
This gives
   \begin{align}
 \Sigma_{jk}(x,s)&=\p_j(x,s|a_k)[\delta_{j,k}+\delta_{j,k+1}]+\frac{1}{2}\Sigma_{jk}(x,s)[\widetilde{f}_{k}^-(a_k^-,s)+\widetilde{f}_k^+(a_k^+,s)]\nonumber \\
 &+\frac{1}{2}\Sigma_{j,k-1}(x,s)\widetilde{f}_{k-1}^+(a_{k}^-,s)+\frac{1}{2}\Sigma_{j,k+1}(x,s)\widetilde{f}_{k+1}^-(a_{k}^+,s)].
 \end{align}

Comparison with equations (\ref{Sigja})--(\ref{Sigjc}) implies that the above equation can be rewritten in the matrix form
\begin{align}
\sum_{l=1}^{m-1} \overline{\Theta}_{kl}(s) \Sigma_{jl}(s)=-\p_j(x,s|a_k)[\delta_{j,k}+\delta_{j,k+1}],
\label{osigp}
\end{align}
where $\overline{\bm \Theta}(s)$ is a tridiagonal matrix with non-zero elements
\begin{subequations}
\begin{align}
\overline{\Theta}_{k,k}(s) &=\overline{d}_k(s)\equiv \widetilde{f}_{k}^-(a_k^-,s)+\widetilde{f}_{k}^+(a_k^+,s)-1,\  k=1,\ldots m-1,\\
\overline{\Theta}_{k,k-1}(s)&=  \overline{c}_{k}(s)\equiv \widetilde{f}_{k-1}^+(a_k^-,s),\quad k=2,\ldots m-1,\\
\overline{\Theta}_{k,k+1}(s)&=\overline{ b}_{k}(s)\equiv \widetilde{f}_{k+1}^-(a_k^+,s),\quad k=1,\ldots,m-2.
\end{align}
\end{subequations}
Assuming that the matrix $\overline{\bm \Theta}(s)$ is invertible, we obtain the formal solution
\begin{equation}
\Sigma_{jk}(x,s)=-\overline{\Theta}^{-1}_{kj}(s) \p_j(x,s|a_j) -\overline{\Theta}^{-1}_{k,j-1}(s) \p_j(x,s|a_{j-1}) .
\end{equation}
Substituting into equation (\ref{Frenewal2}) yields the result
\begin{align}
  \label{fFrenewal2}
  \widetilde{\rho}_j(x,s) &=  \p_j(x,s) +\frac{1}{2}\sum_{k=1}^{m-1}[\overline{\Theta}^{-1}_{kj}(s) \p_j(x,s|a_j) +\overline{\Theta}^{-1}_{k,j-1}(s) \p_j(x,s|a_{j-1})] \\
  &\hspace{3cm}\times  [\widetilde{f}_{k}^-(s)
+\widetilde{f}_{k}^+(s)] .\nonumber
 \end{align}

 \paragraph{Equivalence of first and last renewal equations for exponential killing}
 
 An important check of our analysis is to show that the solution (\ref{fFrenewal2}) of the first renewal equation is equivalent to the solution (\ref{lastsol}) of the last renewal equation when each round of reflecting BM is killed according to an independent exponential distribution for each local time threshold. Since $\p_j(x,s|x_0)$ then satisfies Robin boundary conditions at $x=a_{j-1},a_j$ we find that
\begin{subequations}
\begin{align}
\overline{\Theta}_{k,k}(s) &= \kappa_{k}[\p_{k+1}(a_k,s|a_{k})+\p_k(a_k,s|a_{k})]-1=\Theta_{kk}(s),\\
\overline{\Theta}_{k,k-1}(s)&=  \kappa_{k-1}\p_{k}(a_{k-1},s|a_{k})=  \kappa_{k-1}\p_{k}(a_{k},s|a_{k-1})=\Theta_{k,k-1}(s),\\
\overline{\Theta}_{k,k+1}(s)&=\kappa_{k+1}  \p_{k+1}(a_{k+1},s|a_{k})=\kappa_{k+1}  \p_{k+1}(a_{k},s|a_{k+1})=\Theta_{k,k+1}(s).
\end{align}
\end{subequations}
We have used two important properties of partially reflected BM:
\medskip

\noindent i) Symmetry of the Green's function $\p(x,s|x_0)=\p(x_0,s|x)$.
\medskip

\noindent ii) The solution for the functions $\Sigma_{jk}(x,s)$ is obtained by introducing the initial conditions (\ref{initial}). The FPT densities are thus evaluated at the initial points $a_k^{\pm}$. This means that when we impose the Robin boundary conditions we do not pick up the additional constant term in equations (\ref{kola}) and (\ref{kolb}).
 \medskip
 
\noindent  It follows that the solution  (\ref{fFrenewal2}) reduces to the form
 \begin{align}
 \widetilde{\rho}_j(x,s) &=   \p_j(x,s)+\sum_{k=1}^{m-1} \left ( \p_j(x,s|a_{j-1})\Theta^{-1}_{k,j-1}(s) + \p_j(x,s|a_{j})\Theta^{-1}_{kj}(s) \right )\nonumber \\
 &\times \kappa_k  [\p_k(a_k,s)+\p_{k+1}(a_{k},s)].
\end{align}
Finally, using the fact that
$\kappa_k\Theta^{-1}_{kj}(s) =\kappa_{j}\Theta^{-1}_{jk}(s)$,
we recover the solution (\ref{lastsol}).

\begin{figure}[t!]
  \centering
  \includegraphics[width=13cm]{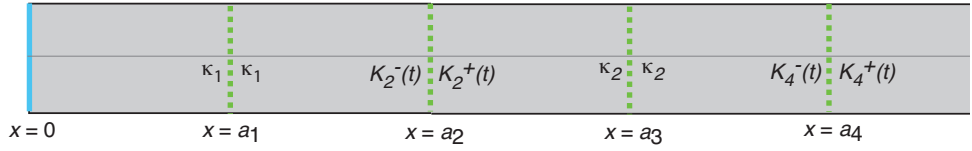}
  \caption{A 1D layered medium partitioned by a sequence of semi-permeable interfaces that alternate between symmetric constant permeabilities $\kappa_j$, $j=1,3,\ldots$ and asymmetric time-dependent permeabilities $K_j^{\pm}(t)$, $j=2,4,\ldots$.}
  \label{fig5}
\end{figure}

\paragraph{Non-exponential killing} The above analysis shows that the same solution structure holds for both exponential and non-exponential killing, provided that we express the tridiagonal matrix $\overline{\Theta}_{ij}$ in terms of the conditional FPT densities $\widetilde{f}_k^{\pm}( a_k^{\pm},s)$, $\widetilde{f}_{k-1}^{+}(a_k^-,s)$ and $\widetilde{f}_{k+1}^{-}(a_k^+,s)$. The latter are themselves determined from equations (\ref{fm}) and (\ref{fp}). One configuration that is analytically tractable is a 1D domain with a sequence of semi-permeable barriers whose distributions $\Psi_j^{\pm}$ alternate between exponential and non-exponential. For example, suppose $\Psi_j^-(\ell)=\e^{-2\kappa_j/D_j}$ and $\Psi_j^+(\ell)=\e^{-2\kappa_j/D_{j+1}}$ for all odd layers $j=1,3,\ldots$, whereas $\Psi_{j}^{\pm}(\ell)$ are non-exponential for even layers $j=2,4,\ldots$. Combining the analysis of the FPT densities in section 5.2 with the analysis of the first renewal equation and its relationship with the last renewal equation, we obtain the following generalization of the interfacial boundary conditions (\ref{LTlayerb}):
\begin{equation}
D_j\partial_x\rho_j(a_j^-,s)=D_{j+1}\partial_x \rho_{j+1}(a_j^+,s) =\frac{1}{2}[\widetilde{K}_j^+(s)\rho_{j+1}(a_j^+,s)-\widetilde{K}_j^-(s)\rho_{j}(a_j^-,s)],
\end{equation}
with $\widetilde{K}^{\pm}_j(s)=2\kappa_j$ for odd $j$ and
\begin{equation}
\widetilde{K}^-_j(s)=\frac{D_j\widetilde{\psi}^-_j(h_j(s))}{\widetilde{\Psi}^-_j(h_j(s))},\quad \widetilde{K}^+_j(s)=\frac{D_{j+1}\widetilde{\psi}^+_j(\overline{h}_{j+1}(s))}{\widetilde{\Psi}^+_j(\overline{h}_{j+1}(s))}
\label{Ker}
\end{equation}
for even $j$, with $h_j(s)$ and $\overline{h}_{j+1}(s)$ given by equations (\ref{hj}) and (\ref{hj2}), respectively. We thus have the setup shown in Fig. \ref{fig5}. Note, in particular, that the time-dependent permeability kernels of the even interfaces are asymmetric.

\paragraph{Permeability kernels for the gamma distribution} For the sake of illustration, suppose that $\psi_j^{\pm}(\ell) $ for even $j$ are given by the gamma distributions
\begin{equation}
\label{psigam}
\psi_j^{\pm}(\ell)=\frac{z_j^{\pm}(z_j^{\pm}\ell)^{\mu-1}\e^{-z_j^{\pm} \ell}}{\Gamma(\mu)}, \mu >0,
\end{equation}
where $\Gamma(\mu)$ is the gamma function. The corresponding Laplace transforms are
\begin{equation}
\widetilde{\psi}_j^{\pm} (z)=\left (\frac{z_j^{\pm}}{z_j^{\pm}+z}\right )^{\mu},\quad \widetilde{\Psi}_j^{\pm}(z)=\frac{1-\widetilde{\psi}_j^{\pm}(z)}{z}.
\end{equation}
If $\mu=1$ then $\psi_j^{\pm}$ reduce to the exponential distributions with constant reactivity $\kappa_j $. The parameter $\mu$ thus characterizes the deviation of $\psi_j^{\pm}(\ell)$ from the exponential case. If $\mu <1$ ($\mu>1$) then $\psi_j^{\pm}(\ell)$ decreases more rapidly (slowly) as a function of the local time $\ell$. Substituting the gamma distributions into equations (\ref{Ker}) yields
 \begin{align}
 \widetilde{K}^-_j(s)&=\frac{D_jh_j(s)(z_j^-)^{\mu}}{[z_j^- +h_j(s)]^{\mu}-(z_j^-)^{\mu}},\ \widetilde{K}^+_j(s)=\frac{D_{j+1}\overline{h}_{j+1}(s)(z_j^+)^{\mu}}{[z_j^+ +\overline{h}_{j+1}(s)]^{\mu}-(z_j^+)^{\mu}}.
 \end{align}
 If $\mu=1$ then $\widetilde{K}^{\pm}_j(s)=2\kappa_j$ as expected. On the other hand if $\mu=2$, say,  then
  \begin{align}
 \widetilde{K}^-_j(s)&=\frac{2\kappa_j}{2+D_jh_j(s)/2\kappa_j },\quad \widetilde{K}^+_j(s)=\frac{2\kappa_j}{2+D_{j+1}\overline{h}_{j+1}(s)/2\kappa_j }.
 \end{align}
 The corresponding time-dependent kernels $K_j^{\pm}(t)$ are normalizable since
 \begin{subequations}
 \begin{align}
 \int_0^{\infty}K_j^{-}(t)dt=\widetilde{K}_j^{-}(0)=\frac{2\kappa_j}{2+\kappa_j^{-1}\kappa_{j-1}/[1+2\kappa_{j-1}L_j /D_j)},\\
 \int_0^{\infty}K_j^{+}(t)dt=\widetilde{K}_j^{+}(0)=\frac{2\kappa_j}{2+\kappa_j^{-1}\kappa_{j+1}/[1+2\kappa_{j+1}L_j /D_{j+1})}.
 \end{align}
 \end{subequations}
 However, the kernels are heavy-tailed with infinite moments. For example,
\begin{align}
\langle t\rangle_{-}&\equiv  \frac{1}{\widetilde{K}_j^{-}(0)}\int_0^{\infty}tK_j^{-}(t)dt=-\frac{1}{\widetilde{K}_j^{-}(0)}\lim_{s\rightarrow 0}\partial_s\widetilde{K}_j^{-}(s)\nonumber \\
&=\frac{1}{\widetilde{K}_j^{-}(0)}\lim_{s\rightarrow  0}\frac{D_jh_j'(s)/2}{[2+D_jh_j(s)/2\kappa_j ]^2}=\frac{D_j}{4\kappa_j}\frac{\widetilde{K}_j^{-}(0)}{2\kappa_j}\lim_{s\rightarrow 0}h_j'(s)=\infty.
\end{align}
  That is, all moments are infinite since all derivatives of $h_j(s)$ are singular at $s=0$.
  An analogous result was previously found for a single interface in 1D and 3D \cite{Bressloff23a,Bressloff23b}.
  
  \section{Discussion} In this paper we developed a probabilistic framework for analyzing single-particle diffusion in heterogeneous multi-layered media. Our approach was based on a multi-layered version of snapping out BM. We showed that the distribution of sample trajectories satisfied
  a last renewal equation that related the full probability density to the probability densities of partially reflected BM in each layer. The renewal equation was solved using a combination of Laplace transforms and transfer matrices. We also proved the equivalence of the renewal equation and the corresponding multi-layered diffusion equation in the case of constant permeabilities. We then used the renewal approach to incorporate a more general probabilistic model of semipermeable interfaces. This involved killing each round of partially reflected BM according to a non-Markovian encounter-based model of absorption at an interface. We constructed a corresponding first renewal equation that related the full probability density to the FPT densities for killing each round of reflected BM. In particular, we showed that non-Markovian models of absorption can generate asymmetric, heavy-tailed  time-dependent permeabilities. 
  
In developing the basic mathematical framework, we focused on relatively simple examples such as identical layers with constant permeabilities or alternating Markovian and non-Markovian interfaces. We also restricted our analysis to the Laplace domain rather than the time domain. However, it is clear that in order to apply the theory more widely, it will be necessary to develop efficient numerical schemes for solving the last or first renewal equations in Laplace space, and then inverting the Laplace transformed probability density to obtain the solution in the time domain. In the case of non-Markovian models of absorption at both ends of a layer, it will also be necessary to compute the double inverse Laplace transform of the local time propagator and evaluate the resulting double integral in equation (\ref{oo}).
Another computational issue is developing an efficient numerical scheme for simulating sample trajectories of snapping out BM in heterogeneous multi-layer media. 

Finally, from a modeling perspective, it would be interesting to identify plausible biophysical mechanisms underlying non-Markovian models of semi-permeable membranes. As previously highlighted within the context of encounter-based models of absorption \cite{Grebenkov20,Grebenkov22,Bressloff22,Bressloff22a}, various surface-based reactions are better modeled in terms of a reactivity that is a function of the local time. 
For example, the surface may become progressively
activated by repeated encounters with a diffusing
particle, or an initially highly reactive surface may become less active due to multiple interactions with the particle (passivation) \cite{Bartholomew01,Filoche08}.

\end{document}